\documentclass[preprint,12pt,3p]{elsarticle}
\usepackage{framed,multirow}
\usepackage{float}
\usepackage{amssymb}
\usepackage{latexsym}
\usepackage{url}
\usepackage{color}
\usepackage{mathrsfs} % mathscript
\definecolor{inchworm}{rgb}{0.7, 0.93, 0.36}

\usepackage{algorithm}
\usepackage{algorithmic}
\usepackage{amsmath,amssymb,amsfonts}
\usepackage{graphicx}
\usepackage{textcomp}
\usepackage{cuted}
\usepackage{dblfloatfix} 
\usepackage{hyperref}

\hypersetup{colorlinks = true,
    linkcolor=blue,
    filecolor=cyan,
    citecolor = blue,
    urlcolor=blue}%color hyperlinks

\usepackage[english]{babel}
\addto\extrasenglish{

} % Autoref Section not section%
\addto\captionsenglish{} % Image caption Figure to Fig.
\usepackage[labelfont=bf]{caption} % Only Fig. in bold and caption in normal text
\usepackage{relsize} %Increase math font size in table

\newcommand\eatpunct[1]{} % Remove full-stop of paragraph
\usepackage{booktabs} % table stylish
\usepackage[bb=boondox]{mathalfa} % double 1
\usepackage{float} % force place figure
\usepackage{makecell} % Wrap table row content
\usepackage{flushend} % Balance references
\usepackage[nameinlink,capitalize]{cleveref} % Equation to Eq. (1)
\usepackage{setspace}

\journal{Applied Soft Computing}

\begin{document}

%%%%%%%%%%%%%%%%%%%% TITLE PAGE %%%%%%%%%%%%%%%%%%%%%%%

\begin{titlepage}
\doublespacing
{\centering{\Large \textbf{An Explainable Contrastive-based Dilated Convolutional Network with Transformer for Pediatric Pneumonia Detection}}\\
Chandravardhan Singh Raghaw, Parth Shirish Bhore, Mohammad Zia Ur Rehman, Nagendra Kumar\\}

\vspace{2em}

{\large Highlights}

\begin{itemize}
\item Contrastive-inspired convolutional network with transformer for pneumonia detection.
\item Combined contrastive-entropy loss to improve feature learning and classifier learning.
\item Feature fusion strategy to capture joint global and fine-grained local information.
\item XCCNet enables deep insights for pneumonia classification through explainability.
\item Extensive evaluations of four datasets demonstrate the superiority of XCCNet.
\end{itemize}

\vspace{2em}

\noindent This is the preprint version of the accepted paper.\\
\noindent This paper is accepted in \textbf{Applied Soft Computing, 2024.}
\\
DOI: \url{https://doi.org/10.1016/j.asoc.2024.112258}
\end{titlepage}
% %%%%%%%%%%%%%%%%%%%%%%%%%%%%%%%%%%%%%%%%%%%%%%%%%%%%%%%

%%%%%%%%%%%%%%%%%%%%%%%%%%%%%%%%%%%%%%%%%%%%%%%%%%%%%%%

% \verso{Raghaw \textit{et~al.}}

\begin{frontmatter}
\title{An Explainable Contrastive-based Dilated Convolutional Network with Transformer for Pediatric Pneumonia Detection}

\author[1]{Chandravardhan Singh Raghaw}
\ead{phd2201101016@iiti.ac.in}

\author[1]{Parth Shirish Bhore}
\ead{cse200001015@iiti.ac.in}

\author[1]{Mohammad Zia Ur Rehman}
\ead{phd2101201005@iiti.ac.in}

\author[1]{Nagendra Kumar\corref{cor1}}
\ead{nagendra@iiti.ac.in}
\cortext[cor1]{Corresponding author}

\address[1]{Department of Computer Science and Engineering, Indian Institute of Technology (IIT) Indore, Indore 453552, India}

\begin{abstract}
% DONE
Pediatric pneumonia remains a significant global threat, posing a larger mortality risk than any other communicable disease. According to UNICEF, it is a leading cause of mortality in children under five and requires prompt diagnosis. Early diagnosis using chest radiographs is the prevalent standard, but limitations include low radiation levels in unprocessed images and data imbalance issues. This necessitates the development of efficient, computer-aided diagnosis techniques. To this end, we propose a novel  E\textbf{X}plainable \textbf{C}ontrastive-based Dilated \textbf{C}onvolutional \textbf{Net}work with Transformer (XCCNet) for pediatric pneumonia detection. XCCNet harnesses the spatial power of dilated convolutions and the global insights from contrastive-based transformers for effective feature refinement. A robust chest X-ray processing module tackles low-intensity radiographs, while adversarial-based data augmentation mitigates the skewed distribution of chest X-rays in the dataset. Furthermore, we actively integrate an explainability approach through feature visualization, directly aligning it with the attention region that pinpoints the presence of pneumonia or normality in radiographs. The efficacy of XCCNet is comprehensively assessed on four publicly available datasets. Extensive performance evaluation demonstrates the superiority of XCCNet compared to state-of-the-art methods.
% We have publicly released the XCCNet code and augmented dataset on GitHub (\url{https://github.com/cvsingh007/xccnet}) to facilitate its use by the research community.
\end{abstract}

\begin{keyword}
%% Keywords
% \KWD \newline
Pneumonia detection \sep
Convolutional neural network \sep
Transformer \sep
Contrastive learning \sep
Explainable AI
\end{keyword}

\end{frontmatter}

%% main text
%%%%%%%%%%%%%%% INTRODUCTION START %%%%%%%%%%%%%%%
\section{Introduction}
\label{sec:introduction}
% DONE

As per the World Health Organization (WHO), pneumonia is the most severe respiratory disease that affects the lungs and creates problems in breathing, and it is caused by bacterial, fungal, and viral infections~\citep{Liang2020}. In the respiratory system of a healthy individual, the lungs consist of minuscule structures known as alveoli that inflate with air during inhalation. However, in the case of pneumonia, the alveoli become engorged with a combination of fluid and pus, leading to respiratory challenges and diminished oxygen uptake. Moreover, pneumonia is the most common disease among children worldwide\footnote{\scriptsize{\url{https://www.unicef.org/health/childhood-diseases}}}. As per the United Nations International Children's Fund (UNICEF), after every 43 seconds, a child loses her life because of pneumonia\footnote{\scriptsize{\url{https://data.unicef.org/topic/child-health/pneumonia/}}}. The death ratio caused by pneumonia per year is more than any other infectious disease. In the year 2019, more than 7,40,180 kids under the age of five died because of pneumonia\footnote{\scriptsize{\url{https://www.who.int/news-room/fact-sheets/detail/pneumonia}}}. As per the joint research investigation of John Hopkins University and Save the Children NGO, the death count from pneumonia will reach 11 million if no suitable measures are taken~\citep{Fernandes2021}. Thus, it is essential to tackle the problem of pediatric pneumonia detection quickly in the early stages.

Timely and accurate detection of pediatric pneumonia is an effective way to save many children's lives. Standard diagnostic techniques for pediatric pneumonia include Computerized Tomography (CT) scans, Magnetic Resonance Imaging (MRI), and chest X-ray (CXR or Chest Radiographs). Due to the painless and non-invasive way of evaluation, CXR is one of the most commonly used radiological tests for the diagnosis of pneumonia and numerous other lung diseases. However, the pneumonia diagnosis using CXR depends on the expertise of radiologists~\citep{Liang2020}. As per the Radiologist Society of North America (RSNA)\footnote{\scriptsize{\url{https://www.rsna.org/news/2022/may/Global-Radiologist-Shortage}}}, there is a global shortage of radiologists, and the situation is worse in developing countries where medical resources are limited. Also, CXR-based diagnosis requires accurate and meticulous observation~\citep{Raoof2012}, which is time-consuming. Furthermore, in some cases, consultation of multiple specialists and more expensive tests are required, which raises the overall cost of diagnosis~\citep{Delrue2011}. 

Providing assistance to specialists in diagnosing medical conditions is crucial, and offering a second opinion can be instrumental in achieving this goal. Artificial Intelligence (AI) algorithms are designed to perform tasks that are conventionally associated with human intelligence~\citep{He2019} to offer the relevant second opinion. Deep Learning, a branch of AI, comprises complex data processing, such as images, through multiple layers of artificial neurons in a network. Hence, the implementation of AI-based Deep Learning methods enhances the productivity of radiologists, brings down medical expenses, and expedites the diagnosing process, which helps in the early detection of pneumonia in children.

Various methodologies have been suggested to tackle the identification of pneumonia by using chest X-ray (CXR). Liu \textit{et al.}~\citep{Liu2024effctm} propose an integrated convolution and transformer-based technique for the classification of pneumonia and normal CXR. Kaya \textit{et al.}~\citep{Kaya2024featurefusion} develop a hierarchical template-matching approach for relevant feature learning. Features are extracted from multiple CNN models with different architectures and classified using majority voting. Kili{\c{c}}arslan \textit{et al.}~\citep{Kilicarslan2023} utilized the novel activation function named SupEx, which is a combination of ReLU and TanhExp activation function. Okolo \textit{et al.}~\citep{Okolo2022} proposed an enhanced Vision Transformer named IEViT for the detection of pediatric pneumonia utilizing convolutions in transformer-based architecture. Most of the widely utilized Deep Learning classifiers, including MobileNet \citep{Kaya2023}, SqueezeNet \citep{Alharbi2022}, VGG-16 \citep{Gupta2022}, VGG-19, Inception-V3 \citep{Pal2023}, ResNet-50, and Xception \citep{Desai2022}, have attained a considerably high level of accuracy \citep{Ayan2022,Trivedi2022}. Chattopadhyay \textit{et al.} developed a deep feature selection method \citep{Chattopadhyay2022}, which uses meta-heuristic feature selection using a sine cosine algorithm for pneumonia diagnosis. However, existing methods primarily emphasize feature extraction techniques, and they often overlook the fundamental challenges associated with extracting features from chest X-rays (CXRs). 
% In contrast, XCCNet tackles these challenges head-on by focusing not only on feature extraction but also on enhancing CXRs for improved feature quality and mitigating data skewness. This comprehensive approach leads to superior results.

\begin{figure*}[!ht]
    \centering
    \includegraphics[width=\textwidth]{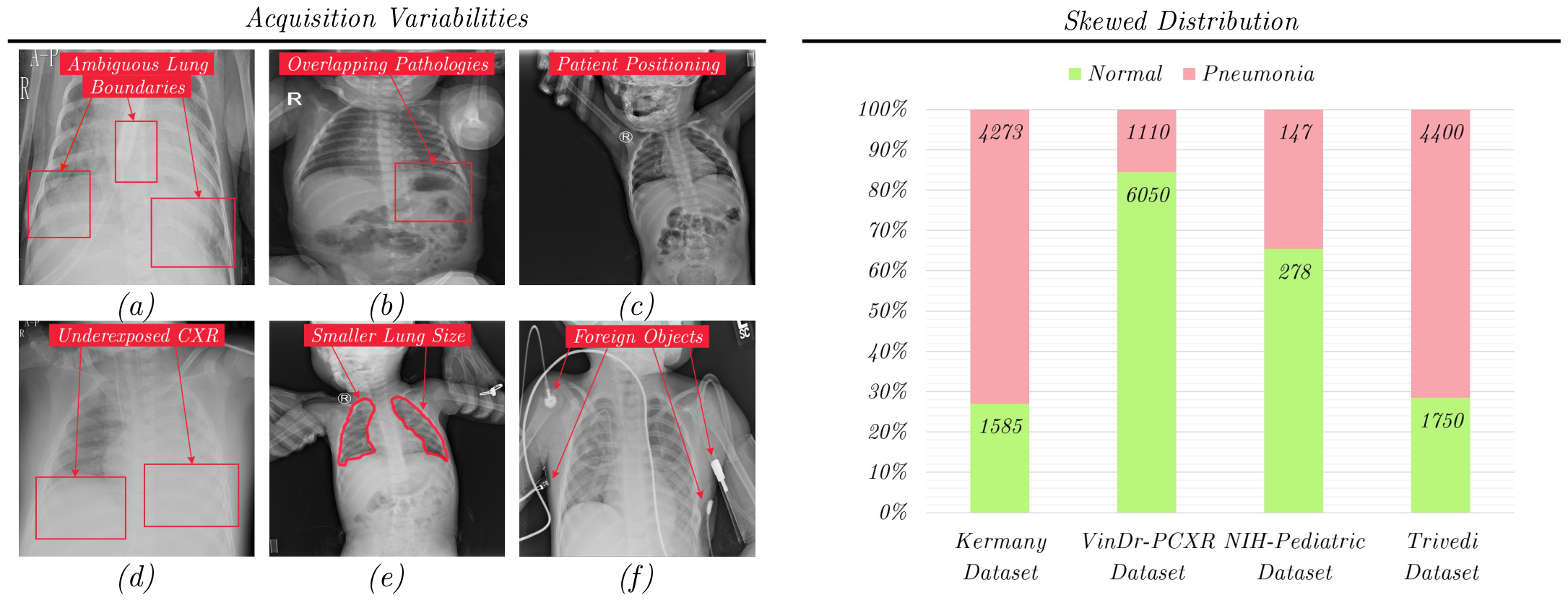}    
    \caption {Left: Visualizing Anatomical and Acquisition Variabilities in chest X-Rays labeled (a) to (f), Right: Highlighting Skewed Distribution of Chest X-Rays in the Pneumonia Dataset}
    \label{fig:challenges}
\end{figure*}

Deep learning models hold immense potential for pediatric pneumonia diagnosis, yet they face several challenges. One major obstacle lies in leveraging low-radiation, unprocessed chest X-rays. These images present complexities due to the inherent variability of pediatric lungs, including ambiguous boundaries, overlapping organs, and inconsistent patient positioning (as depicted in~\autoref{fig:challenges}). This variability, coupled with the presence of foreign objects and reduced lung size, further complicates the analysis. An additional hurdle arises from the skewed distribution of pediatric CXR data, where one class (e.g., pneumonia) has significantly fewer samples compared to the other (normal). This class imbalance can mislead models, causing them to favor the majority class and miss the minority class (pneumonia). Consequently, models struggle to learn both local and global patterns within the X-rays, hindering their diagnostic accuracy. Another challenge lies in accurately identifying infected regions within the X-ray. Precisely pinpointing these areas significantly improves feature extraction and, consequently, the model's overall performance. Inaccurate identification of the Region of Interest (ROI) leads to the model learning irrelevant patterns from the X-ray. Segmenting the infected area allows the model to focus on key features that differentiate normal and pneumonia cases. Finally, capturing the diverse features present in chest X-rays poses another significant challenge. Features extracted from a single extractor might not encapsulate the full spectrum of relevant information. This challenge can be mitigated by fusing features derived from multiple extractors, resulting in a more comprehensive and robust representation of the input data. By addressing these challenges, deep learning models can be effectively leveraged for accurate and reliable pediatric pneumonia detection.

\subsection*{Contributions of the Proposed Work}
% DONE
This research introduces a novel Explainable Contrastive-based Dilated Convolutional Network with Transformer (XCCNet) framework to tackle the limitations faced by existing methods in pediatric pneumonia diagnosis. XCCNet leverages the combined power of contrastive learning, transformers, and convolutional networks to effectively capture both global and spatial features from chest X-rays (CXRs). Contrastive learning serves as a loss function, bridging the gap between global transformer features and fine-grained spatial features from the convolutional network. This approach enhances CXR representation by emphasizing intra-class similarity and inter-class dissimilarity. Furthermore, XCCNet incorporates a robust preprocessing module that amplifies raw CXRs into enhanced CXRs, followed by lung area segmentation for precise feature extraction. To address the skewed distribution of CXR data, the framework utilizes a synthetic data generator, creating high-resolution synthetic CXRs that effectively mitigate the impact of data imbalance, a common challenge in medical datasets.

\begin{itemize}
    \item We propose an Explainable Contrastive-based Dilated Convolutional Network with Transformer (XCCNet) for pediatric pneumonia detection. XCCNet leverages the synergy of contrastive learning, transformers, and convolutional networks to efficiently capture both global and spatial features from chest X-rays (CXRs).
    \item XCCNet employs contrastive learning as a combined loss function, bridging the gap between global and spatial features. This boosts CXR representation by focusing on intra-class similarity and inter-class dissimilarity.
    \item We introduce a chest X-ray Preprocessing Module that transforms low-radiation, unprocessed CXRs into enhanced CXRs, minimizing anatomical variabilities and focusing on pneumonia-related features.
    \item We present an adversarial-based data augmentation unit that utilizes adversarial learning principles to generate high-resolution synthetic CXRs and balance the skewed data distribution.
    \item To enhance interpretability, we integrate an explainability module that generates attention maps. These maps visualize the framework's focus during diagnosis, assigning weights to regions that significantly impact pediatric pneumonia detection.
    \item XCCNet surpasses fourteen state-of-the-art pneumonia detection methods on four publicly available datasets. This highlights its potential for clinical applications.
\end{itemize}

This article is organized into distinct sections, as follows. \autoref{sec:related} provides an in-depth examination of existing literature on the detection of pediatric pneumonia. \autoref{sec:method} provides an elaborate account of the proposed methodology. \autoref{sec:expeval} provides an overview of the evaluation process conducted during the experiments. \autoref{sec:discussion} discusses the importance of the proposed framework and outlines the potential future research directions. Finally, \autoref{sec:conclusion} summarizes the essential findings of the proposed framework.

%%%%%%%%%%%%%%% INTRODUCTION END %%%%%%%%%%%%%%%

%%%%%%%%%%%%%%% RELATED WORK START %%%%%%%%%%%%%%%
\section{Related Work}
\label{sec:related}
This section presents a comprehensive overview of existing research on pediatric pneumonia detection using chest X-rays (CXRs). To facilitate understanding, we categorize these techniques into four distinct groups: Convolutional Neural Network-based Pneumonia Detection, Transformer-based Pneumonia Detection, Hybrid Model-based Pneumonia Detection, and Explainable-based Pneumonia Detection methods.

\subsection{Convolutional-based Pneumonia Detection}
\label{sec:rwcnn}
Convolutional Neural Network (CNN)-based approaches for pneumonia detection leverage the power of convolutional layers. These layers employ filters to scan input chest X-ray (CXR) data and capture local patterns and structures, ultimately aiding in pneumonia classification~\citep{Han2024dmcnn,Rajeashwari2024deepcnn}. Several recent advancements have been made in this area, Yi \textit{et al.}~\citep{Yi2023dl} introduced a scalable and interpretable Deep Convolutional Neural Network (DCNN) specifically designed to extract primary features from CXR images for pneumonia identification. Kili{\c{c}}arslan \textit{et al.}~\citep{Kilicarslan2023} developed a novel activation function called SupEx, which combines the properties of TanhExp and ReLU algebraically while preserving crucial differentiability, thereby influencing subsequent layers within the CNN during backpropagation. Trivedi \textit{et al.}~\citep{Trivedi2022} proposed a lightweight deep learning architecture built on MobileNet. This architecture primarily utilizes depth-wise separable convolutions alongside GlobalAveragePooling2D for automated pneumonia detection.

CNN-based techniques have shown advancements in pneumonia detection, and they often struggle to capture both intricate local details and long-range dependencies within CXRs. This limitation can lead to misclassification, particularly for subtle pneumonia cases. XCCNet addresses these limitations by combining the strengths of CNNs and transformers. It leverages CNNs for efficient local feature extraction and employs a transformer-based module for capturing crucial long-range dependencies. This complementary approach empowers XCCNet to achieve superior performance in pediatric pneumonia detection compared to traditional CNN-based methods.

\subsection{Transformer-based Pneumonia Detection}
\label{sec:rwtransformer}

Transformer-based Pneumonia Detection harnesses the power of the self-attention mechanism to capture long-range dependencies and global relationships within chest X-ray (CXR) images for improved pneumonia classification. This technique allows the model to learn complex interdependencies between different image regions, potentially beyond the capabilities of traditional CNNs~\citep{Hao2024dbmvit,Chen2024intercnn}. Several recent studies showcase the application of this approach, Usman \textit{et al.}~\citep{Usman2022} utilized a pre-trained vision transformer on both the CheXpert dataset and a pediatric pneumonia dataset. By employing random weight initialization, they aim to facilitate learning long-range features from CXR images. Okolo \textit{et al.}~\citep{Okolo2022} proposed an integrated approach combining features extracted by a ResNet-based skip connection with those extracted by a vision transformer. This parallel integration enables their model to simultaneously learn local and global information within CXR images for pneumonia classification. Singh \textit{et al.}~\citep{Singh2024evit} emphasize the efficiency of vision transformers in capturing global features for pneumonia detection in CXR images. Their work further highlights the potential of this approach in this domain.

Transformer-based methods demonstrate proficiency in capturing long-range dependencies within chest X-rays (CXRs), their efficacy in extracting fine-grained local features, a critical aspect for accurate pneumonia diagnosis, can be limited. This constraint restricts their ability to differentiate subtle pneumonia cases from normal CXRs effectively. XCCNet, conversely, transcends these limitations by strategically incorporating a contrastive loss function between convolutional neural networks (CNNs) and transformers. This combined loss function optimizes the overall feature representation by leveraging CNNs for robust local feature extraction while concurrently incorporating a dedicated transformer module to capture global dependencies. This synergistic approach empowers XCCNet to achieve superior performance in pediatric pneumonia detection tasks on chest X-rays.

\subsection{Hybrid Model-based Pneumonia Detection}
\label{sec:rwhybrid}

In the domain of pneumonia detection, hybrid deep learning methodologies integrate the strengths of pre-trained architectures, including Convolutional Neural Networks (CNN) and transformers. This synergistic approach leverages the combined capabilities of these models to achieve superior performance~\citep{Mann2024hybrid,Gupta2023hybridcnn,raghaw2024cotconet}. Several recent studies showcase diverse implementations of this strategy, Ayan \textit{et al.}~\citep{Ayan2022} proposed an automatic pediatric pneumonia diagnosis system utilizing an ensemble of convolutional neural networks (CNNs) including Xception, ResNet-50, and MobileNet. To preserve spatial features, they incorporate a global average pooling layer. Chattopadhyay \textit{et al.}~\citep{Chattopadhyay2022} introduced a feature selection framework using the Sine Cosine Algorithm (SCA). This framework leverages a pre-trained DenseNet-201 for feature extraction, followed by SCA-based selection to improve pneumonia detection performance. Prakash \textit{et al.}~\citep{Prakash2023tl} presented stacked ensemble learning on features extracted from ResNet50V2, ResNet101V2, ResNet152V2, Xception, and DenseNet169. Additionally, they utilize Kernel PCA for dimensionality reduction and a stacking classifier combining XGBClassifier, Support Vector Classifier, Logistic Regression, Nu-SVC, and K-Nearest Neighbor. Finally, a Nu-SVC meta-classifier performs the final classification. Kaya \textit{et al.}~\citep{Kaya2024featurefusion} developed a hierarchical template-matching approach for relevant feature learning. They extract features from multiple CNN models with different architectures and combine them using majority voting for classification. Lie \textit{et al.}~\citep{Liu2024effctm} introduced a hierarchical mechanism that combines a CNN, Transformer, and Multi-Layer Perceptron. This approach integrates spatial and global feature extraction for improved pneumonia detection.

Hybrid architectures integrating convolutional neural networks (CNNs) and transformers have shown promise in boosting performance, and they can introduce potential redundancy during feature extraction. This redundancy can complicate the interpretation of model decisions and potentially hinder the generalizability of the model to unseen data. XCCNet addresses these limitations by employing a feature fusion module. These fused features subsequently promote interpretability by enabling the analysis of distinct feature extraction steps. This targeted approach not only achieves superior performance in pediatric pneumonia detection but also enhances interpretability.

\subsection{Explainable-based Pneumonia Detection}
\label{sec:rwxai}
Explainable Artificial Intelligence (XAI) empowers medical experts by offering visual interpretation tools for trained deep-learning models in pediatric pneumonia detection~\citep{Dai2024xai,Hasan2023fpcnn}. This capability facilitates focused analysis on the most relevant image regions, aiding in informed diagnostic decision-making. Several recent studies showcase the application of XAI in this domain, Ukwuoma \textit{et al.}~\citep{Ukwuoma2023xet}, proposed an explainable hybrid model combining deep ensemble techniques (GoogleNet, DenseNet201, VGG16) with a transformer encoder. This integration generates explainable maps that aid in understanding the model's predictions for accurate CXR-based pneumonia detection. Yang \textit{et al.}~\citep{Yang2022xai} presented an explainable deep learning approach that removes lung backgrounds and utilizes GradCAM to generate explainability maps from a fine-tuned VGG16 model. This allows users to understand which regions contribute most to the prediction. Hroub \textit{et al.}~\citep{Hroub2024xdl} demonstrated the effectiveness of efficient data augmentation with InceptionV3, exceeding the performance of vision transformers. They further visualize their technique's decision-making process using various tools like AblationCAM, EigenGradCAM, GradCam++, RandomCAM, and GradCAM.

Explainable Artificial Intelligence (XAI) methods offer valuable insights into the decision-making processes of deep learning models for pediatric pneumonia detection, their dependence on specific architectures, and potential limitations in providing comprehensive interpretability. In contrast, XCCNet addresses these challenges by adopting a comprehensive explainability framework that leverages various attention heatmaps, including Grad-CAM, Grad-CAM++, Score-CAM, and LIME. These heatmaps visually highlight the primary areas of feature learning where the model focuses its attention. This targeted approach empowers XCCNet to achieve superior interpretability and accuracy compared to existing techniques for pediatric pneumonia detection on chest X-rays.

%%%%%%%%%%%%%%% RELATED WORK END %%%%%%%%%%%%%%%

%%%%%%%%%%%%%%% METHODOLOGY START %%%%%%%%%%%%%%%

\section{Methodology}
\label{sec:method}
% DONE
The initial section of this paper, \autoref{sec:problem}, comprehensively outlines the problem formulation and provides a preliminary overview of the proposed XCCNet. Subsequently, \autoref{sec:xccnetmethod} explores its internal components. The Chest X-ray Preprocessing module is introduced in \autoref{sec:cxpu}. This is followed by the Spatial Feature Extraction module~\autoref{sec:sfx}, which extracts local fine-grained spatial features from the preprocessed images. Next, \autoref{sec:cotfx} introduces the Contrastive-based Transformer Feature Extractor module, which leverages contrastive learning for refining the extracted features. Finally, \autoref{sec:dhffc} describes the Dual Hybrid Feature Fusion and Classification module, which combines the local and global features and classifies them as normal or pneumonia.

\subsection{Problem Formalization and Framework Overview}
\label{sec:problem}
% DONE

We aim to detect pediatric pneumonia in chest X-rays by precisely classifying them as normal or pneumonia. The objective is to process the input $x_{raw}$ and predict the pneumonia with the output, $\mathcal{P} \in$ [\textit{normal. pneumonia}]. Let us consider $XCCNet_{train} = \biggr\{\Bigr(x_{raw}, \mathcal{P}\Bigr)\biggr\}^N_{i=1}$ to be the set of $N$ chest X-ray samples. With $XCCNet_{train}$, the objective to predict the output class for input image $f: x_{raw} \gets \mathcal{P}$. The $XCCNet_{train}$ model predicts $\mathcal{P}$ as $\mathcal{\tilde{P}}$, and the difference contributes to the combined loss $\mathscr{L}_{combined}\Bigr[\mathcal{\tilde{P}}:\mathcal{P}\Bigr]$, where $\mathscr{L}_{combined}$ is the aggregate loss of the Spatial Feature Extractor (SFx) module and Contrastive-based Transformer Feature Extractor (CoTFx) Module.

\begin{figure}[!ht]
  \centering
  \includegraphics[width=0.6\linewidth]{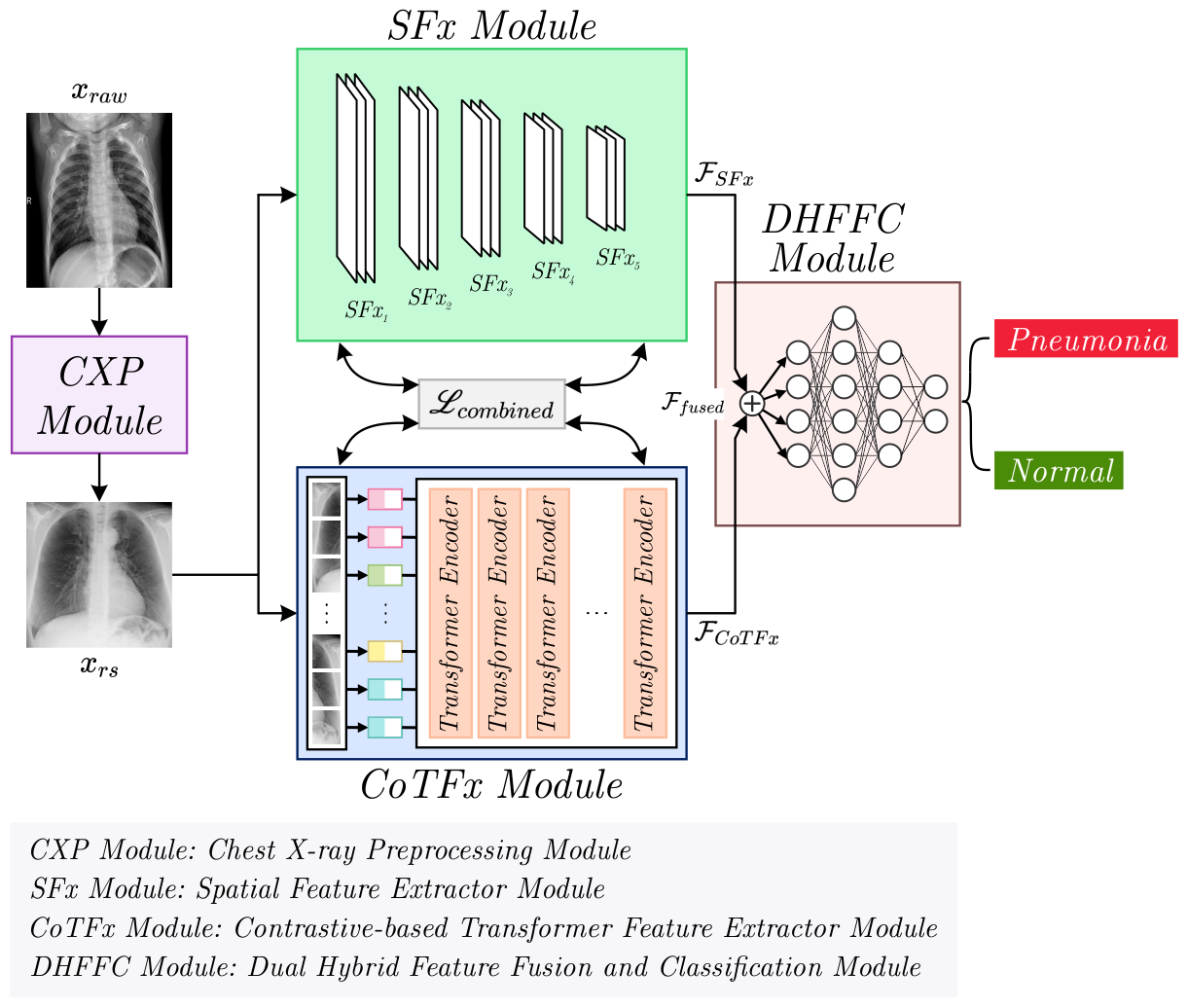}
  \caption{The workflow of our proposed approach for detecting pediatric pneumonia in chest X-rays.  The process begins with the input raw CXR $x_{raw}$, which the Chest X-ray Preprocessing Module it into enhanced CXR enhanced CXR $x_{rs}$. The Spatial Feature Extractor Module captures fine-grained features from the enhanced CXR. These features are then fused with global features extracted by the Contrastive-based Transformer Feature Extractor Module within the Dual Hybrid Feature Fusion and Classification Module. Finally, this module classifies the fused features as normal CXR or pneumonic CXR.}
  \label{fig:xccnetoverview}
\end{figure}

\autoref{fig:xccnetoverview} details the architectural overview of the XCCNet framework, designed for detecting pediatric pneumonia from chest X-rays (CXRs).  XCCNet comprises four integral modules: (a) a chest X-ray preprocessing module, which refines raw CXRs to improve feature extraction, aiding both spatial and contrastive-based transformer feature extraction. It reduces data skewness, amplifies low-radiation CXRs, extracts lung information, and selectively reduces rib visibility while preserving internal soft tissue details; (b) a spatial feature extraction module utilizes dilated convolutions to capture fine-grained spatial features from CXRs, such as shape, color, texture, and hierarchical information. Dilated convolutions enlarge the receptive field while maintaining feature map resolution, allowing for the extraction of richer spatial details; (c) a contrastive-based transformer feature extractor module combines a transformer architecture to explore global contextual dependencies with contrastive learning. The contrastive learning loss function promotes feature learning by focusing on intra-class similarity and inter-class dissimilarity, leading to improved feature discrimination. Additionally, a cross-entropy loss is integrated to guide the network toward accurate classification based on the learned features. Finally, (d) a dual hybrid feature fusion and classification module fuses spatial and global features with the same dimensions along the channel dimension. The fused features are fed into a deep classifier to categorize the input CXR as normal or pneumonia.

\begin{figure*}[!ht]
    \centering
    \includegraphics[width=0.9\textwidth]{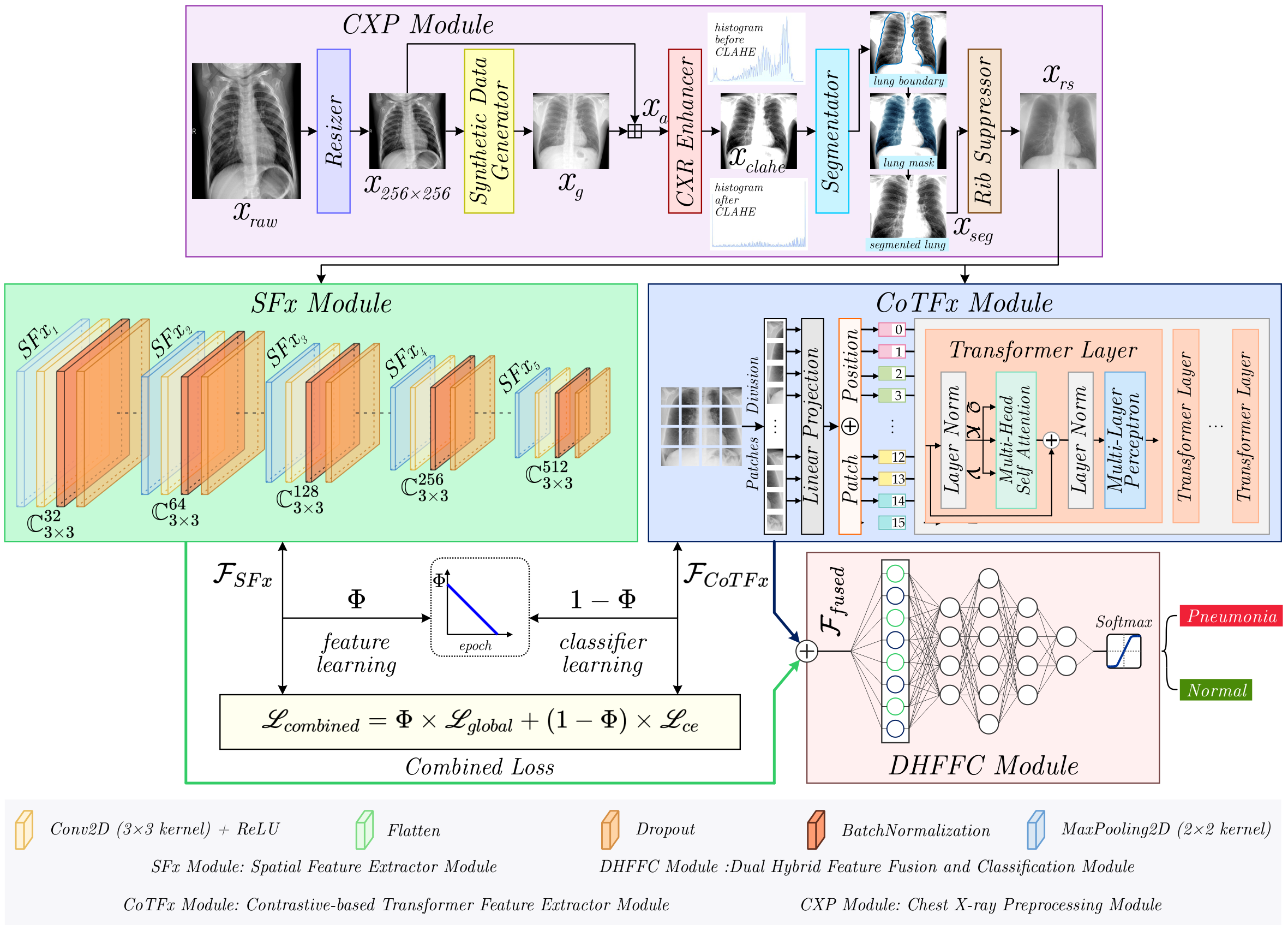}    
    \caption {A detailed breakdown of the Explainable Contrastive-based Dilated Convolutional Network with Transformer (XCCNet) framework for pediatric pneumonia detection in chest X-rays (CXRs). The process begins with the Chest X-ray Preprocessing Module (CXP), which transforms the raw input into an enhanced CXR ($x_{raw} \rightarrow x_{rs}$) to enhance feature quality. The enhanced CXR $x_{rs}$ is then fed into both the Spatial Feature Extractor Module (SFx) and the Contrastive-based Transformer Feature Extractor Module (CoTFx) in parallel. SFx captures fine-grained spatial features $\mathcal{F}_{SFx}$, while CoTFx extracts global features $\mathcal{F}_{CoTFx}$. A combined loss function $\mathscr{L}_{combined}$ guides the network towards accurate classification. Finally, the features are fused $\mathcal{F}_{SFx} \oplus \mathcal{F}_{CoTFx} \rightarrow \mathcal{F}_{fused}$ and used to categorize the CXR as normal or pneumonia.}
    \label{fig:xccnet}
\end{figure*}

\subsection{Explainable Contrastive-based Dilated Convolutional Network with Transformer}
\label{sec:xccnetmethod}
Explainable Contrastive-based Dilated Convolutional Network with Transformer (XCCNet) is composed of four primary modules, namely the Chest X-ray Preprocessing (CXP) module, Spatial Feature Extraction (SFx) Module,  Contrastive-based Transformer Feature Extractor (CoTFx) module, Dual Hybrid Feature Fusion and Classification (DHFFC) module. These modules are the foundation pillars of XCCNet, each having its unique role, including CXR preprocessing to final pneumonia classification. In the next sections, we will explore the internal architectural details of each module. \autoref{fig:xccnet} illustrated the detailed architecture of the XCCNet framework.

\subsubsection{Chest X-ray Preprocessing Module}
\label{sec:cxpu}
This section details various image processing techniques employed within the Chest X-ray Preprocessing (CXP) module to prepare input Chest X-rays (CXR). The CXP module acts as a critical gateway, transforming raw CXRs through a series of targeted refinements: (a) Resizing, (b) Synthetic Data Generation, (c) Image Enhancing, (d) Segmentation, and (e) Rib Suppression. These combined processing steps work synergistically to enhance low-level features within the CXRs. This ultimately facilitates improved extraction and maximizes the information captured from each sample, leading to more accurate and robust analysis. \autoref{fig:xccnet} outlines the pipeline of the CXP module.

\paragraph[]{Resizer: \eatpunct}

We resize the collected CXR to spatial dimensions of $256 \times 256$ while preserving image quality. The resizing mechanism utilizes bicubic interpolation. Bicubic interpolation stands out as a favored method for image resizing due to its efficient balance between computational cost and quality enhancement. The bicubic interpolation in line 4 can be mathematically expressed in \cref{eq:bicubic}:

\begin{equation}
    \begin{aligned}
    \label{eq:bicubic}
        interpolate_{bicubic} \left(x_{raw}, \frac{x_n}{\alpha_x}, \frac{y_n}{\alpha_y} \right)
        = \sum_{x} \sum_{y} B(x - u)\, B(y - v)\,I(x_o + u, y_o + v)
    \end{aligned}    
\end{equation}

Here, $x_o = \left\lfloor \frac{x_n}{\alpha_x} \right \rfloor$ and $y_o = \left\lfloor \frac{y_n}{\alpha_v} \right \rfloor$ are the corresponding integer coordinates in the original image $x_{raw}$. Variable $u$ and $v$ iterate over the neighboring pixel offsets (-1, 0, 1). $B (\cdot)$ is the bicubic basis function evaluated at $(\cdot)$ and $I(x_o + u, y_o + v)$ is the intensity value at the neighboring pixel.

\paragraph[]{Synthetic Data Generator: \eatpunct}
In medical imaging, scarcity of images poses a significant challenge to effective model training, often hindering convergence. Data augmentation techniques address this challenge by generating synthetic data mimicking real-world samples. Classical data augmentation techniques include geometric transformations such as flipping, rotation, scaling, and skewing. While conventional data augmentation relies on simple replication techniques, these methods may not adequately capture the full spectrum of clinical characteristics present in complex medical images despite enhancing dataset diversity to some extent. However, for medical images with complex imaging textures, the samples obtained by this method may not meet the clinical characterization of the images. This requires an alternative augmentation technique that generates synthetic medical samples with similar textures and patterns observed in medical scans. In response to this, we devised a novel Adversarial-based Data Augmentation (ADA) unit for the generation of high-resolution synthetic CXRs. Recent research in medical imaging demonstrates a significant increase in the utilization of data augmentation methods~\citep{Chen2022ganreview, Xu2024cdagan}.

\begin{figure}[!ht]
  \centering
  \includegraphics[width=0.6\linewidth]{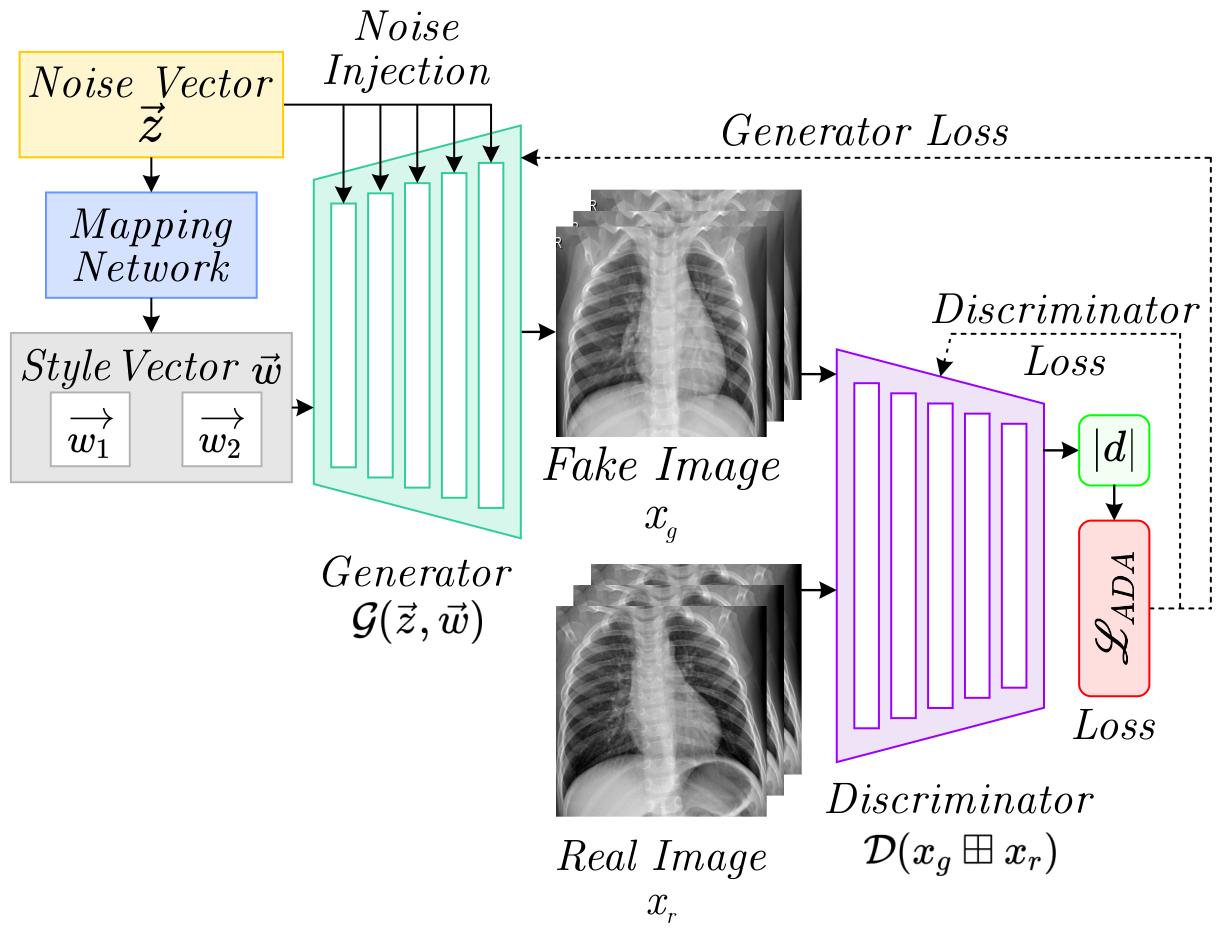}
  \caption{ The figure presents the proposed Adversarial-based Data Augmentation (ADA) unit, designed to generate synthetic chest X-rays. This unit incorporates a noise vector $\overrightarrow{z}$, a style network $\overrightarrow{w}$, a Generator $\mathcal{G}$, and the discriminator $\mathcal{D}$. Specifically, the Generator $\mathcal{G}$ utilizes the noise vector $\overrightarrow{z}$ and the style vector $\overrightarrow{w}$ to generate synthetic CXRs, while the discriminator $\mathcal{D}$ evaluates a combination of real and synthetic CXRs. The ADA unit utilizes a generator-adversarial loss function $\mathscr{L}_{ADA}$ to guide the training process. This function ensures that the generated synthetic CXRs effectively mimic the characteristics of real chest X-rays, thereby mitigating data skewness and enhancing the training process.}
  \label{fig:adaunit}
\end{figure}

\subparagraph{Adversarial-based Data Augmentation Unit:}
We aim to design and implement an Adversarial-based Data Augmentation (ADA) Unit for medical image analysis tasks. ADA unit will leverage adversarial learning principles to synthetically generate additional data, mitigating the impact of data imbalance issues in medical datasets. \autoref{fig:adaunit} illustrates the architecture of the ADA unit to generate synthetic CXR.

\autoref{fig:adaunit} illustrates the Adversarial-based Data Augmentation (ADA) unit. A noise vector $\overrightarrow{z}$, where $\overrightarrow{z} \in \left\{\overrightarrow{z_1}, \ldots ,\overrightarrow{z_n} \right\}$ sampled from a normal distribution, serves as input to the mapping network. The network employs a Multi-Layer Perceptron (MLP) with numerous hidden units to transform $\overrightarrow{z}$ into a a style vector $\overrightarrow{w}$. For synthesis, the style vector bifurcates into $\overrightarrow{w_1}$ and $\overrightarrow{w_2}$ to inject noise into separate upper and lower portions of the synthetic image. The generator block $\mathcal{G}(\overrightarrow{z}$,$\overrightarrow{w})$ then takes these vectors as inputs to produce fake images $x_g$, given in~\cref{eq:6}.

\begin{equation}
    \label{eq:6}
    x_g = \mathcal{G}(\overrightarrow{z},\overrightarrow{w})
\end{equation}

The generator block utilizes stacked Adaptive Instance Normalization (AdaIN) to normalize the feature map of synthetic images with $\overrightarrow{w}$, as mathematically presented in in~\cref{eq:7}:

\begin{equation}
    \label{eq:7}
    AdaIN\Bigr(x_g, \bigr(\overrightarrow{w}\bigr)\Bigr) = \sigma{(\overrightarrow{w})} \cdot
    \frac{\Bigr(x_g - \mu \bigr(x_g\bigr)\Bigr)}{\sigma(x)} + \mu(\overrightarrow{w})
\end{equation}

Here, $\sigma(\overrightarrow{w})$ and $\mu(\overrightarrow{w})$ are the standard deviation and mean of $\overrightarrow{w}$, respectively, and $\mu(x_g)$ and $\sigma(x_g)$ represent the mean and standard deviation of the feature maps of $x_g$. The generated $x_g$ and real image $x_r \in x_{raw}$ are simultaneously permuted (denoted as $\boxplus$) and fed into the discriminator $\mathcal{D}(x_g\boxplus x_r)$, which outputs a scalar value $|d|$ indicating whether the input is real or fake. The discriminator's and generator's objective functions are given in \cref{eq:8} and ~\cref{eq:9}, respectively.

\begin{equation}
    \label{eq:8}
    \mathscr{L}_{\mathcal{D}} = -[log(\mathcal{D}(x_g \uplus x_r))] - E[log(1-\mathcal{D}(G(\overrightarrow{z}, \overrightarrow{w})]
\end{equation}

\begin{equation}
    \label{eq:9}
    \mathscr{L}_{\mathcal{G}} = E[log(\mathcal{D}(\mathcal{G}(\overrightarrow{z}, \overrightarrow{w})]
\end{equation}

The ADA unit learns parameters from both the generator and discriminator, minimizing the total loss given in~\cref{eq:10}.

\begin{equation}
    \label{eq:10}
    \mathscr{L}_{ADA} = \mathscr{L}_{\mathcal{D}} + \mathscr{L}_{\mathcal{G}} + \lambda \cdot \mathscr{L}_{reg}
\end{equation}

This loss function comprises three components: $\mathscr{L}_{\mathcal{D}}$ measures differences between fake and real images, $\mathscr{L}_{\mathcal{G}}$ encourages realistic $x_g$, and $\mathscr{L}_{reg}$ enforces a unit vector constraint on $\overrightarrow{w})$, where $\lambda$ is a constant. Balancing these components leads to the generation of more realistic synthetic images, $x_g$. Furthermore, the generated synthetic $x_g$ is combined with real CXR $x_r$,  leading to an augmented dataset $\textbf{X}_\textbf{a}$, where $x_a \in \textbf{X}_\textbf{a}$.

\paragraph[]{CXR Enhancer: \eatpunct}
To enhance image contrast, we employed Contrast Limited Adaptive Histogram Equalization (CLAHE). CLAHE demonstrates advantages over Adaptive Histogram Equalization (AHE), particularly for improving low-contrast images in the medical domain~\citep{Shamrat2023clahe, Rifai2024imgenhance}. In our experiments, we empirically selected a window size of $8 \times 8$ to employ the HE in each local region (window) of the image. Additionally, a clipping scale of $2.0$ (represented as $CLAHE 2.0$) was chosen, resulting in more excellent contrast enhancement compared to $CLAHE 1.0$. Compared to $CLAHE 1.0$ with fixed clipping, $CLAHE 2.0$ utilizes adaptive clipping limits, mitigating artifacts and preserving image fidelity. To validate this, we computed the correlation coefficient between the enhanced CXR images and the original CXR images in the dataset for various CLAHE scales. Our findings reveal that the correlation coefficient for CLAHE 2.0 surpasses other scales, leading us to select it for contrast enhancement with an $8 \times 8$ window and generate the final enhanced image $x_{clahe}$.

\paragraph[]{Segmentator: \eatpunct}

\begin{figure*}[!ht]
    \centering    
    \includegraphics[width=0.8\textwidth]{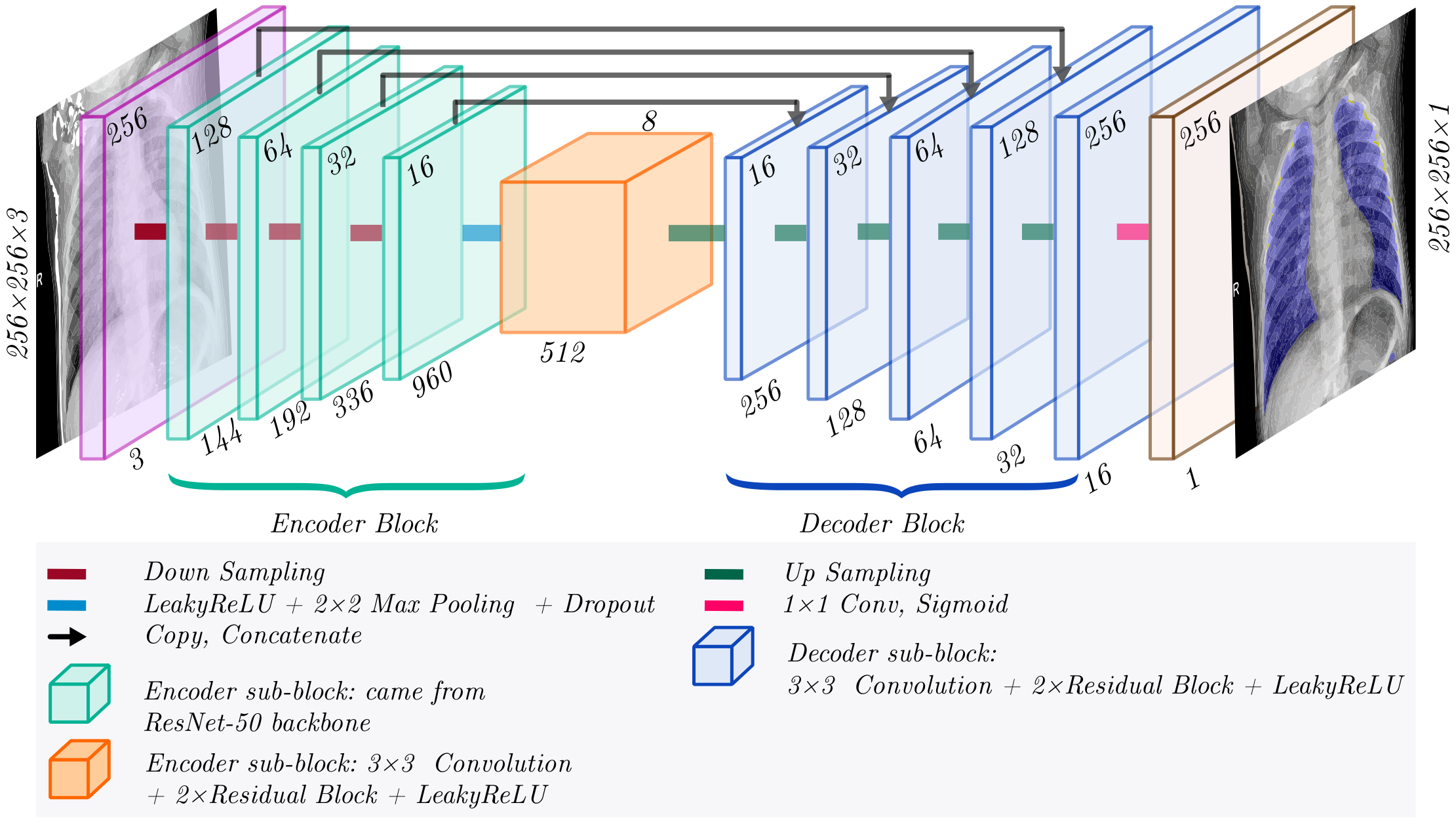}
    \caption {The ResUNet++ architecture for lung segmentation employs a pre-trained ResNet-50 model for feature extraction within the encoder. Transposed convolutions upsample the features, and a final convolutional layer with sigmoid activation generates the lung segmentation mask}
    \label{fig:segmentator}
\end{figure*}

Lung segmentation is a vital step in pneumonia classification, facilitating the isolation of affected regions for enhanced analysis. This refined Region Of Interest (ROI) enables more precise feature extraction and subsequent classification, leading to superior results compared to analyzing the entire chest X-ray. However, achieving robust segmentation remains challenging due to factors such as image variability, overlapping structures, and subtle intensity differences. Therefore, developing a robust lung segmentation technique is essential for efficiently extracting lung regions from Chest X-rays (CXRs) and enabling accurate subsequent analysis tasks. To this end, we propose a modified ResUNet++ architecture for lung segmentation. \autoref{fig:segmentator} represents the architecture of modified ResUNet++. ResUNet++ augments the UNet architecture with skip connections, enabling the decoder to refine segmentation maps by incorporating information from earlier encoder layers. Additionally, ResUNet++ incorporates dense connections between all subsequent layers, facilitating information flow and feature reuse for enhanced performance.

\autoref{fig:segmentator} depicts the depicts the encoder block, which utilizes ResNet50 as its backbone network. This backbone employs convolutional layers with residual connections for enhanced information flow. The encoder function, represented by $\mathscr{E}_{seg}$, takes an input-enhanced CXR and generates a feature map $\vec{\mathcal{F}}_{seg}$ containing spatial features. This $\vec{\mathcal{F}}_{seg}$  information is shared across all encoder and decoder layers through skip connections.

The decoder block function, $\mathscr{D}_{seg}\left(\vec{\mathcal{F}}_{seg}\right)$, leverages the feature map $\vec{\mathcal{F}}_{seg}$ generated by the encoder block to construct a segmentation map. This map identifies the lung boundary and creates the binary mask $\mathcal{B}_{mask}$ for input CXR, assigning a value of $1$ to pixels on the boundary and $0$ otherwise. To minimize the error between the predicted $\mathcal{B}_{mask}$ and the ground truth segmentation map $\mathcal{S}_{map}$, the binary cross-entropy loss function, $\mathscr{L}_{seg} \Bigr( \mathcal{S}_{map}, \mathcal{B}_{mask} \Bigr)$, is employed as defined in~\cref{eq:segloss}.

\begin{equation}
    \begin{aligned}
    \label{eq:segloss}
    \mathscr{L}_{seg}
    \Bigr( \mathcal{S}_{map}, \mathcal{B}_{mask} \Bigr)=\\
     &\hspace{-25mm}-\Bigr[ \mathcal{S}_{map} \times log\bigr( \mathcal{B}_{mask} \bigr) 
    +\bigr(1 - \mathcal{S}_{map}\bigr)
    \times log\bigr(1 - \mathcal{B}_{mask}\bigr)\Bigr]\\
    \end{aligned}
\end{equation}

This loss function serves as a quantitative measure of the overall segmentation prediction accuracy. Finally, the generated mask $\mathcal{B}_{mask}$ is overlaid onto the input image using a logical $AND$ ($\wedge$) operation. This yields the final lung mask $\mathcal{L}_{mask}$, which is then cropped based on mask pixels to obtain the segmented CXR image $x_{seg}$.

\paragraph[]{Rib Suppressor: \eatpunct}
Chest X-rays (CXR) offer valuable diagnostic information on both soft tissues and bones, which is crucial for radiologists and clinicians in diagnosis and treatment decisions. However, overlapping structures like clavicles and ribs can hinder accurate analysis. To address this challenge, rib suppression techniques actively aim to reduce the visibility of bony structures, particularly ribs, while preserving internal soft tissue details. This refined CXR facilitates improved diagnostic inference and accurate analysis.

\begin{figure*}[!ht]
  \centering
  \includegraphics[width=\textwidth]{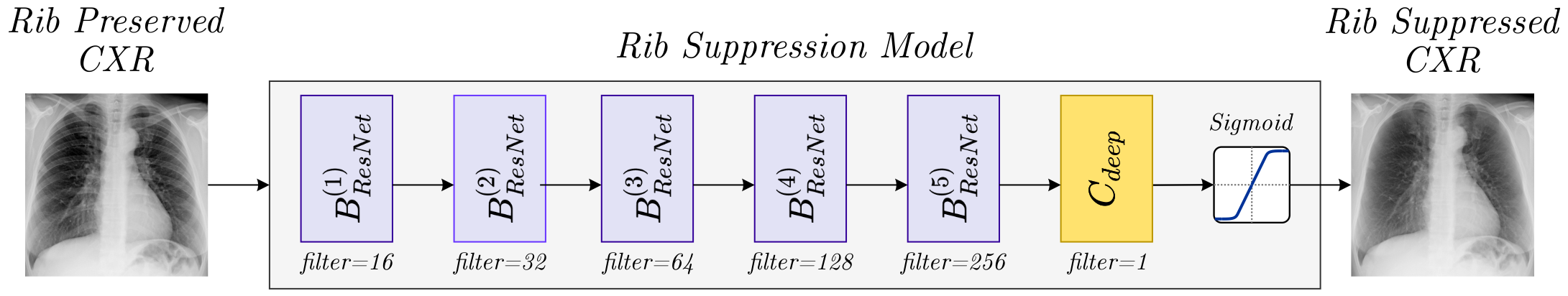}
  \caption{The proposed $RS_{ResNet}$ unit consists of five residual blocks $(B_{ResNet})$, each utilizing $3 \times 3$ kernels and zero-padding to maintain the original dimensions of the input. The deepest convolutional layer $C_{deep}$, followed by a sigmoid activation function, generates the rib-suppressed CXR.}
  \label{fig:ribsup}
\end{figure*}

Our experiments leverage the $RS_{ResNet}$ unit, a modified ResNet-based rib suppression model inspired by inspired by Rajaraman \textit{et al.}~\citep{Rajaraman2021}. $RS_{ResNet}$ actively removes bones while preserving soft tissue details, leading to improved visibility of internal organs. The architecture of $RS_{ResNet}$ is presented in~\autoref{fig:ribsup}.

The $RS_{ResNet}$ unit is a sequential model comprising five stacked ResNet blocks $B_{ResNet}^{(i)}$ for $i=1$ to $5$. Each block employs $16$, $32$, $64$, $128$, and $256$ filters respectively, utilizing $3 \times 3$ kernels with zero padding to maintain image dimensions. These blocks leverage skip connections, effectively mitigating vanishing gradient issues and enabling the reuse of activations from previous layers until adjacent layer weights are updated.  Finally, the deepest convolutional layer $(C_{deep})$, featuring a single filter with zero padding and a sigmoid activation function, generates the final rib-suppressed image $x_{rs}$. \cref{alg:cxpu} provides a step-by-step explanation of the Chest X-ray Preprocessing (CXP) Module.

% \begin{figure*} % Full-width Algorithm
% \centering{
% \begin{minipage}{0.9\linewidth}
  \begin{algorithm}
    \caption{\textsc{CXP Module}}
    \label{alg:cxpu}
    \begin{algorithmic}
        \vspace{1mm}
        \STATE
        \textbf{Input:} $\langle x_{raw} \rangle \in$ input CXR dimensions\\
        \vspace{0.8mm}
        \textbf{Output:} $\langle x_{rs} \rangle \in$ Rib Suppressed CXR\\
        
        \vspace{1mm}    
        \textbf{procedure:} {$resize_{bicubic}$} $\left( x_{raw}, x_{256 \times 256} \right)$
    \end{algorithmic}
    \begin{algorithmic}[1]
    \STATE Calculate scaling factor: $\alpha_x, \alpha_y, \leftarrow \frac{256}{I_w}, \frac{256}{I_h}$
    \STATE Initialize an empty resized image ${h_n \times w_n}$ \\
    \FOR {each output pixel $(x_n,y_n)$}
        \STATE $x_{256 \times 256} (x_n,y_n)\leftarrow interpolate_{bicubic} (x_{raw}, \frac{x_n}{\alpha_x}, \frac{y_n}{\alpha_y})$
      \ENDFOR \\
      \STATE\textbf{return} The resized CXR $x_{256 \times 256}$
    
    \vspace{1mm}    
    \hspace{-2.75mm}\textbf{procedure:}  {$ADA\, Unit$} $(x_{256 \times 256}, x_a)$
    \STATE Randomly sample a noise vector $(\vec{z})$,and generate style vector $\vec{w}$ using $MLP$ on $(\vec{z})$
    \STATE Generate a synthetic image $x_g$, using generator network $\mathcal{G}$ with inputs $(\vec{z})$ and $(\vec{w})$
    \STATE Select real image $x_r$, calculate the discriminator loss $\mathscr{L}_{\mathcal{D}}$ and generator loss $\mathscr{L}_{\mathcal{G}}$ and update the combined loss $\mathscr{L}_{\mathcal{ADA}}$ with regularization
    \STATE The generated synthetic $x_g$ is combined with real CXR $x_r$ to create augmented CXR $x_a$
    \STATE\textbf{return} The augmented CXR $x_a$
    
    \vspace{1mm}
    \hspace{-2.75mm}\textbf{procedure:}  {$CXR_{enhance}$} $(x_g, x_{clahe})$
    \STATE Apply CLAHE with an $8 \times 8$ window size and clipping scale of 2.0 to enhance the contrast.
    \STATE\textbf{return} The contrast-enhanced CXR $x_{clahe}$

    \vspace{1mm}
    \hspace{-2.75mm}\textbf{procedure:}  {$ResUnet\!+\!+$} $(x_{clahe}, x_{seg})$
    \STATE Encoder $\mathscr{E}_{seg}(x_{clahe})$ generates feature map $\vec{\mathcal{F}}_{seg}\left(x_{clahe}\right)$ 
    \STATE Decoder $\mathscr{D}_{seg}\left(\vec{\mathcal{F}}_{seg}(x_{clahe})\right)$ generate binary mask $\mathcal{B}_{mask}$
    \STATE Lung mask $\mathcal{L}_{mask}$ is obtained by applying logical AND between $\mathcal{B}_{mask}$ and original CXR.
    \STATE Extract pixels within the mask area to obtain segmented CXR.
    \STATE\textbf{return} The segmented CXR $x_{seg}$

    \vspace{1mm}
    \hspace{-2.75mm}\textbf{procedure:}  {$RS_{ResNet}$} $(x_{seg}, x_{rs})$
    \STATE Feature Extraction from input segmented CXR: \\ $\mathcal{F}_{rs} = {\displaystyle \prod_{i=1}^{5}} B_{ResNet}^{(i)} (x_{seg}) $
    \STATE Final rib suppression from extracted features: \\$x_{rs} = C_{deep} (\mathcal{F}_{rs})$
    \STATE\textbf{return} The rib-suppressed CXR $x_{rs}$
    
    \end{algorithmic}    
  \end{algorithm}
% \end{minipage}}
% \end{figure*}

% \subsubsection{Contrastive-based Dilated Convolutional Feature Extraction Module}
% \label{sec:cdcfx}

% Contrastive-based Dilated Convolutional Feature Extraction (CDCFx) module architecture, depicted in \textbf{FIGURE}, leverages two core feature extractor units: the Spatial Feature Extractor (SFx) module and the Contrastive-based Transformer Feature Extractor (CoTFx) module. SFx module focuses on extracting fine-grained spatial features through dilated convolutions, while the CoTFx module utilizes a transformer architecture to explore global contextual dependencies. To enhance consistency in representing features belonging to the same class, contrast learning is employed as a loss function, operating on attention maps generated by the CoTFx module. This unique dual-feature extractor design aims to combine local and global features from both modules effectively. The subsequent section explores the detailed architecture of SFx and CoTFx modules.

% \paragraph[]{Spatial Feature Extractor Module: \eatpunct}
\subsubsection{Spatial Feature Extractor Module}
\label{sec:sfx}
\begin{figure}[!ht]
    \centering    
    \includegraphics[width=0.6\linewidth]{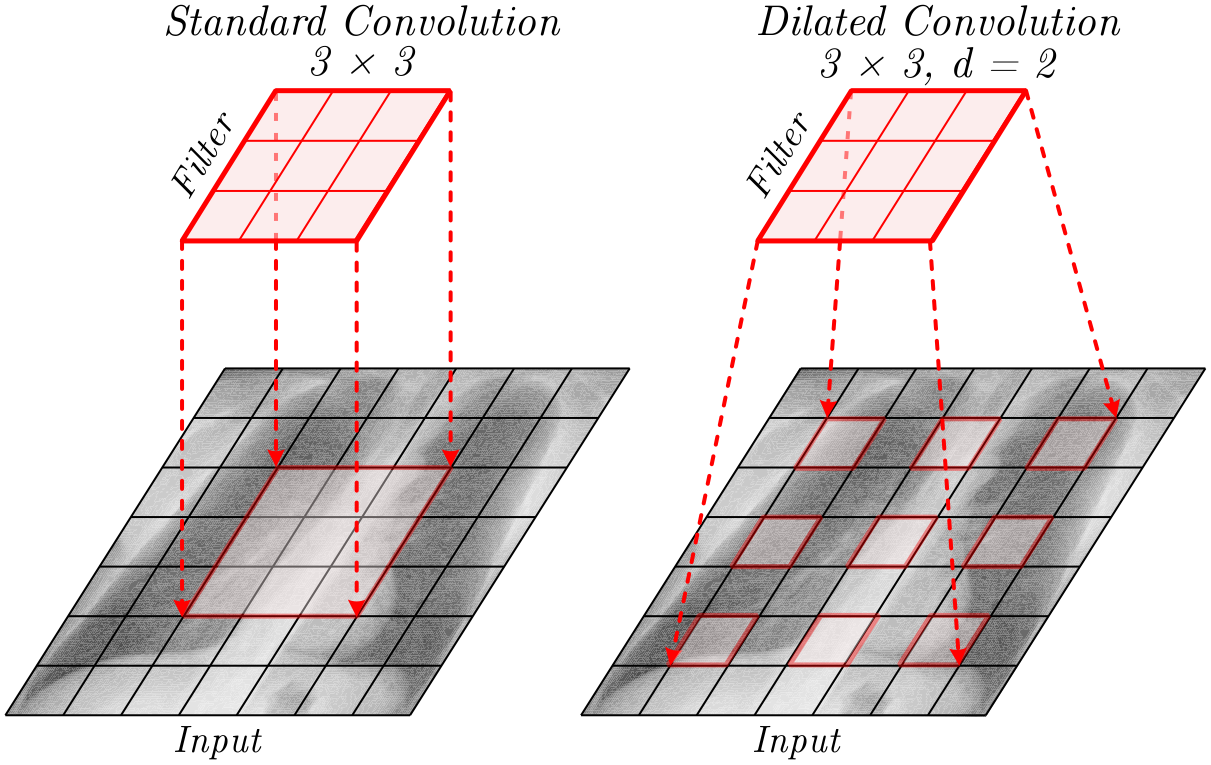}
    \caption {Compared to standard $3 \times 3$ convolutions, dilated convolutions (kernel size $3 \times 3$, dilation rate 2) capture a larger receptive field, enabling the extraction of broader information.}
    \label{fig:dilatedcnn}
\end{figure}

The Spatial Feature Extractor (SFx) module leverages a dilated convolution network to extract fine-grained spatial features like shape, target color, texture, and hierarchical information from CXR images effectively. Unlike standard convolutions, dilated convolutions allow for a larger receptive field while maintaining output size by inserting gaps between kernel elements.
The functionality of the SFx module is outlined in~\cref{alg:cdcfx}. In the SFx module, a dilation rate of $2$ is applied to a $3 \times 3$ kernel, resulting in a larger receptive field that captures more data while maintaining computational efficiency. \autoref{fig:dilatedcnn} illustrates the difference between standard and dilated convolution. Both have the same kernel size, but dilated convolution utilizes a larger receptive area, preserving resolution in feature maps.

As shown in the \autoref{fig:dilatedcnn}, a dilated convolution with a dilation rate of $2$ and a $3 \times 3$ kernel is functionally similar to a standard convolution with a $5 \times 5$ kernel. \cref{eq:dcnn} formalizes the Dilated Convolution operation:

\begin{equation}
    \label{eq:dcnn}
    y[i, j] = \sum_{m = 0}^{k - 1} \sum_{n = 0}^{k - 1} x[i - d \cdot m, j - d \cdot n] \cdot f[m, n]
\end{equation}

Here, $y[i, j]$ signifies the output value at the spatial location $(i, j)$ within the feature map. The input feature map itself is denoted by $x$ while $f$ represents the filter kernel. Notably, the dimensions of the filter kernel are captured by $k$. The dilation rate, denoted by $d$, controls the spacing between elements within the filter kernel, influencing the receptive field of the convolution operation. Furthermore, the time complexity of dilated convolution $O\biggr(\Bigr(k+\bigr(k-1\bigr) \cdot \bigr(d-1\bigr)\Bigr)^2 \cdot m \cdot n \cdot f\biggr)$ is lower than standard convolution $O\bigr(k^2 \cdot m \cdot n \cdot f\bigr)$, where $k$ is the kernel size, $d$ is the dilation rate, $m$ and $n$ are input image dimensions, and $f$ is the number of filters.

\autoref{fig:xccnet}  illustrates the architecture of the SFx module for feature extraction. Additionally, it incorporates pixel distribution within the image space as part of its feature set. Internally, the SFx module comprises five feature extractor blocks, denoted as $S\!F\!x_i,\, i \in [1,2 \ldots 5]$ each containing a sequence of convolution operations denoted as $\mathbb{C}_{k \times k}^{f}$, where $\mathbb{C}$ represents the Conv2D operation, $k$ represents the kernel size, and $f \in [32,64,128,256,512]$ represents the number of filters employed by each block. To enhance feature robustness, each block incorporates a dropout layer $(\delta_{0.1})$ with a dropout rate of $0.1$, followed by batch normalization $(\beta_n)$ and max-pooling $(\rho_{max})$. Finally, all extracted features are flattened using the flatten layer $(\digamma)$.

The grid-structured input layer directly feeds the rib-suppressed CXR $(x_{rs})$ into a sequence of five Spatial Feature Extractor $(S\!F\!x_i)$ blocks, where $i$ ranges from $1$ to $5$. Each block employs $\mathbb{C}_{k \times k}^{f}$ convolutional operations with ReLU activation to actively extract dominant local features. Notably, ReLU is chosen over Sigmoid and Tanh for its reduced computational cost and prevention of vanishing gradients. Simultaneously, a batch normalization layer $(\beta_n)$ calculates and normalizes the mean and variance of the mini-batch, facilitating faster learning. The normalized output is downsampled by half using a max-pooling layer $(\underset{2\times 2}{\rho_{max}})$ with a window of $2\times 2$ and stride of $2$, retaining key features. To prevent overfitting, a dropout layer $\delta_{0.1}$ layer randomly drops neurons during training, encouraging the network to learn more robust features and reduce dependence on individual neurons. Finally, the flatten layer transforms the multidimensional feature map into a one-dimensional vector, denoted as $\mathcal{F}_{SFx} \in \mathbb{N}^{2304}$ with dimensions $N\times 2304$, where $N$ represents the number of input samples. This $\mathcal{F}_{SFx}$ vector is then fed into the Dual Hybrid Feature Fusion and Classification (DHFFC) module to combine local fine-grained features with long-range global contrastive features.

% \paragraph[]{Contrastive-based Transformer Feature Extractor Module: \eatpunct}
\subsubsection{Contrastive-based Transformer Feature Extractor Module}
\label{sec:cotfx}

% NTR
Contrastive-based Transformer Feature Extractor (CoTFx) module combines the Vision Transformer (ViT) and contrastive learning. The architecture of the ViT is schematically illustrated in \autoref{fig:xccnet}. The network is composed of Patch Division, Position Embedding, and Transformer blocks that focus on extracting global information from the input images. Contrastive learning is introduced as a loss function between features from the SFx module and CoTFx module to boost the CXR representations. Contrastive loss improves feature learning by focusing on intra-class similarity and inter-class dissimilarity. Additionally, a cross-entropy loss is incorporated after contrastive loss to drive the network toward accurate classification based on learned features. Leverages both aspects, potentially leading to better performance than using either loss alone. \cref{alg:cdcfx} outlines the working of the CoTFx module.

\subparagraph{Transformer-based Feature Extractor:}
Transformer-based Feature Extractor leverages the power of Contrastive Language-Image Pretraining (CLIP) framework \citep{Radford2021}. The CLIP model utilizes both image and text encoders to generate semantically similar and dissimilar text-image pairs. In our experimentation, we leverage the visual encoder of the CLIP as a backbone to extract global features. Specifically, the ViT-B/32 variant, as its robust feature representation, is pre-trained by vision-language contrastive learning. The ViT/B-32 captures the long-range dependencies between different CXR regions, potentially leading to improved performance on the classification of normal or pneumonia. The ViT-B/32 image encoder operates in three key stages:

\vspace{1.5mm}
\noindent\textit{(i) Patch Embedding and Positional Encoding: }
The input rib-suppressed CXRs are partitioned into non-overlapping square patches of $16 \times 16$ pixels. Each patch is flattened into a vector and embedded into a lower-dimensional space using a linear layer. This creates a patch embedding $e_i \in \mathbb{R}^{d_p}$ for each patch $i$, where $d_p$ is the embedding dimension. Additionally, positional encoding, represented by $E_{pos}(i)$ is added to each patch embedding to account for their relative location within the image, $\tilde{p}_i = p_i + E_{pos}(i)$. 

\vspace{1.5mm}
\noindent\textit{(ii) Transformer Encoder: }
The core of the ViT-B/32 backbone lies in its 12 stacked transformer encoder blocks, each containing multi-head attention, feed-forward networks, layer normalization, and residual connections. Multi-head attention allows the model to learn long-range global relationships between different patches by attending to their individual embeddings and aggregating relevant information. This can be expressed in~\cref{eq:multihead} and~\cref{eq:head}

\begin{equation}
    \label{eq:multihead}
   MultiHead = Concat \,(head_1, \ldots , head_n)   
\end{equation}

\begin{equation}
    \label{eq:head}
    head_i = Softmax\left( \frac{\mathcal{Q}\mathcal{K}^T_i}{\sqrt{d_{\mathcal{K}}}} \right) \mathcal{V}_i
\end{equation}

Here, $\mathcal{Q}, \, \mathcal{K}, \, \mathcal{V}$ represent the query, key, and value matrices, $n$ signifies the number of attention heads respectively, and $d_{\mathcal{K}}$ is the dimension of the key vectors. In Feed-forward networks provide non-linearity and enhance the network's expressive power. Layer normalization and residual connections ensure stable training and information flow through the network effectively. Layer normalization scales and centers the activations of each layer, while residual connections add the original patch embedding to the transformed output, facilitating gradient flow.

\vspace{1.5mm}
\noindent\textit{(iii) Multi-Layer Perceptron: }
Finally, the output represents the extracted global visual features of the rib-suppressed CXR. The extracted features undergo dimensionality reduction via a Multi-Layer Perceptron (MLP) head, resulting in a compressed 512-dimensional embedding $\mathcal{F}_{CoTFx} \in \mathbb{N}^{512}$ that captures the salient lung patterns, suggesting its potential for improved CXR analysis.

\subparagraph{Contrastive Learning:}
Contrastive learning focuses on a distance metric, aiming to minimize the distance between similar images (positive pairs) and maximize the distance between dissimilar ones(negative pairs)~\citep{Wang2022cltranscnn}. The goal is to narrow the distance between the set of normal and pneumonia samples using the cosine similarity metric. The distance between similar pairs and dissimilar pairs focuses on key distinguishing features generated from SFx module and CoTFx module. Let these features be denoted as $\langle \overrightarrow{x_{S\!F\!x}}, \overrightarrow{x_{C\!o\!T\!F\!x}} \rangle$. The goal is to increase the similarity index for similar features and decrease the similarity index for dissimilar features. The similarity index $\ddot{s}$ is given in~\cref{eq:cosim} is the dot product (i.e. cosine similarity) for vector representation of $\overrightarrow{x_n}$ and $\overrightarrow{x_p}$, and can be mathematically calculated as:

\begin{equation}
    \label{eq:cosim}
    \ddot{s}\, (\overrightarrow{x_{S\!F\!x}}, \overrightarrow{x_{C\!o\!T\!F\!x}}) = \frac{\overrightarrow{x_{S\!F\!x}} \cdot \overrightarrow{x_{C\!o\!T\!F\!x}}}{||\overrightarrow{x_{S\!F\!x}}|| \, ||\overrightarrow{x_{C\!o\!T\!F\!x}}||}
\end{equation}

Here, $\overrightarrow{x_{S\!F\!x}}$ and $\overrightarrow{x_{C\!o\!T\!F\!x}}$ represent the feature representation for SFx module and CoTFx module. $(\cdot)$ represents the dot product of the between two vectors.  A value of $1$ represents more similar vectors, while a values closer to $0$ represent more dissimilar vectors. Next, the features with high similarity index $\ddot{s}$ are taken, and compute the pair-wise contrastive loss between them, as given in~\cref{eq:conloss}

\begin{equation}
    \label{eq:conloss}
    \mathscr{L}_{CL}(x,y) = 
    -\log \left( \dfrac{\exp\left( \dfrac{\ddot{s}\, (\overrightarrow{x_{S\!F\!x}}, \overrightarrow{x_{C\!o\!T\!F\!x}})}{\tau} \right) } 
    {\sum_{k=1}^{2N} \mathbb{1}_{[k\neq x]} 
    \exp \left( \dfrac{\ddot{s}((\overrightarrow{x_{S\!F\!x}}, \overrightarrow{x_{C\!o\!T\!F\!x}})}{\tau} \right)} \right)
\end{equation}

Here, $\tau$ is the temperature parameter that determines the smoothness of the distribution, $N$ is the total number of samples, and $\mathbb{1}_{[k\neq x]} \in \{0,1\}$ is an indicator function results in $1$ iff $k\neq x$. The $\mathscr{L}_{global}$ supports the model for adjacent mapping of similar feature vectors and distant mapping for dissimilar feature vectors. The global contrastive loss $\mathscr{L}_{global}$ is the average of the $\mathscr{L}_{CL}(x,y)$ given in~\cref{eq:lossglobal}.

\begin{equation}
    \label{eq:lossglobal}
    \mathscr{L}_{global} = \frac{1}{N(N-1)}\sum_{x=1}^{N}\sum_{y=1,y\neq x}^{N}
    \mathscr{L}_{CL}(x,y)
\end{equation}

After obtaining the feature representations from the contrastive loss step, we use these features to compute the cross-entropy loss $\mathscr{L}_{CE}$ for pneumonia detection, formulated in~\cref{eq:lossce}:

\begin{equation}
    \label{eq:lossce}
    \mathscr{L}_{ce} = -\frac{1}{N} \sum_{i=1}^{N} \left[ y_i \log(p_i) + (1 - y_i) \log(1 - p_i) \right]
\end{equation}

Here, $\mathscr{L}_{ce}$ represents the cross-entropy loss, $N$ is the number of training samples, and $i$ iterates over each sample. The division by $N$ normalizes the loss across the entire dataset. $y_i$ is the true label for sample $i$ ($0$ for normal, $1$ for pneumonia) and $p_i$ is the predicted probability of pneumonia for sample $i$. The negative sign ensures the loss is minimized when the predicted probability aligns with the true label. Due to the different natures of the two loss functions, feature learning and classifier learning branches have different data sampling strategies. By combining contrastive loss and cross-entropy loss, the network learns discriminative representations through contrastive learning while optimizing for accurate classification of positive pneumonia samples using cross-entropy loss. The combined loss $\mathscr{L}_{combined}$ is the weighted combination of the contrastive loss and the cross-entropy loss, as formulated in~\cref{eq:losscombine}:

\begin{equation}
    \label{eq:losscombine}
    \mathscr{L}_{combined} = \Phi \times \mathscr{L}_{global} + (1- \Phi) \times \mathscr{L}_{ce}
\end{equation}

The variable $\Phi$ is a weighted hyperparameter inversely proportional to the epoch number. $\Phi$ controls the trade-off between the contrastive loss and the cross-entropy loss. Next, the Dual Hybrid Feature Fusion and Classification (DHFFC) Unit fuses the features obtained from the SFx module and CoTFx module and classifies them as normal or pneumonia.

% \paragraph[]{Dual Hybrid Feature Fusion and Classification Unit: \eatpunct}
\subsubsection{Dual Hybrid Feature Fusion and Classification Module}
\label{sec:dhffc}

Previous studies indicate that a combination of transformer and CNN-based techniques extract richer feature information for medical image classification~\citep{Liu2024effctm}. Motivated by the combination of transformer-based techniques with convolutional networks, we present a Dual Hybrid Feature Fusion and Classification (DHFFC) Unit. DHFFC efficiently utilizes the local fine-grained features of the SFx module and the long-range global features from the CoTFx module and fuses them all together to focus on the global context information along with the local detail information. The operation of the DHFFC module is illustrated in~\cref{alg:cdcfx}.

% \begin{figure*} % Full-width Algorithm
% \centering{
% \begin{minipage}{0.9\linewidth}
  \begin{algorithm}
    \caption{\textsc{Feature Extractor and Classification Module}}
    \label{alg:cdcfx}
    \begin{algorithmic}
        \vspace{1mm}
        \STATE
        \textbf{Input:} $\langle x_{rs} \rangle \in$ input Rib Suppressed CXR\\ 
        \vspace{0.8mm}
        \textbf{Output:} $\langle \mathcal{P} \rangle \in$ [normal, pneumonia] \\
        
        \vspace{1mm}    
        \textbf{procedure:} {\textit{SFx}} $\left( x_{rs}, \mathcal{F}_{SFx} \right)$
    \end{algorithmic}
    \begin{algorithmic}[1]
    \STATE Initialize with input: $x_{S\!F\!x} \leftarrow x_{rs}$   
    \FOR {each $S\!F\!x_i$ block, $i \in [1,2 \dots 5]$}
        \STATE $\mathcal{F}_{S\!F\!x} \leftarrow  \delta_{0.1} \biggl(\underset{2\times 2}{\rho_{max}} \Bigr( \beta_n \bigl( \mathbb{C}_{3 \times 3}^f \bigr) \Bigl) \biggr)$
      \ENDFOR \\
    \STATE Transform the multidimensional feature map into a one-dimensional vector: $\mathcal{F}_{SFx} \leftarrow$ \textit{Flatten}$(\mathcal{F}_{SFx})$
    \STATE\textbf{return} 1D feature vector $\mathcal{F}_{SFx} \in \mathbb{N}^{2304}$
    
    \vspace{1mm}    
    \hspace{-2.75mm}\textbf{procedure:}  {\textit{CoTFx}} $(x_{rs}, \mathcal{F}_{CoTFx})$
    \STATE Divide $x_{rs}$ into $16 \times 16$ non-overlapped patches and add positional encoding
    \STATE Process patch embeddings through 12 stacked transformer blocks, with each block using Multi-Head attention to capture long-range dependencies between patches
    \STATE Multi-Layer Perceptron projects the final output $\mathcal{F}_{CoTFx}$    
    \STATE Calculate the global contrastive loss $\mathscr{L}_{global}$ between $\mathcal{F}_{CoTFx}$ and $\mathcal{F}_{SFx}$ followed by cross-entropy loss $\mathscr{L}_{ce}$
    \STATE Calculate the weighted combined loss $\mathscr{L}_{combined}$\\
    $\mathscr{L}_{combined} = \Phi \times \mathscr{L}_{global} + (1- \Phi) \times \mathscr{L}_{ce}$ and minimize them by network training    
    \STATE \textbf{return} The final feature vector $\mathcal{F}_{CoTFx} \in \mathbb{N}^{2304}$
    
    \vspace{1mm}
    \hspace{-2.75mm}\textbf{procedure:}  {\textit{DHFFC}} $(\mathcal{F}_{SFx}, \mathcal{F}_{CoTFx})$
    \STATE Concate channel-wise $\mathcal{F}_{SFx}$ and $\mathcal{F}_{CoTFx}$ features to generate $\mathcal{F}_{fused}$
    \STATE Classify $\mathcal{F}_{fused}$ features into pneumonia or normal CXR using deep classifier
    \STATE\textbf{return} The input CXR $\in$ [normal, pneumonia]    
    \end{algorithmic}    
  \end{algorithm}
% \end{minipage}}
% \end{figure*}

DHFFC utilizes spatial features $\mathcal{F}_{SFx}^{2304}$ and global features $\mathcal{F}_{CoTFx}^{512}$ with the same spatial dimensions to fuse the features along the channel dimension. Feature fusion is mathematically outlined in the following~\cref{eq:wtfusion}, where $\mathcal{F}_{fused} \in \mathbb{N}^{2816}$ and $\oplus$ represents the final fused features and channel-wise concatenation.

\begin{equation}
    \label{eq:wtfusion}
    \mathcal{F}_{fused} = [\mathcal{F}_{SFx}^{2304} \oplus \mathcal{F}_{CoTFx}^{512}]
\end{equation}

Finally, the fused features are fed into the deep classifier with two fully connected layers followed by a sigmoid activation function. The final classification outcome determines whether input CXR is normal or pneumonia.

%%%%%%%%%%%%%%% METHODOLOGY END %%%%%%%%%%%%%%%

%%%%%%%%%%%%%%% EXPERIMENTAL EVALUATIONS START %%%%%%%%%%%%%%%
\section{Experimental Evaluations}
\label{sec:expeval}

Within this section, we present the details of the diverse datasets employed for experimental evaluation, where we directly benchmark our proposed framework against established baselines using rigorous metrics. This performance assessment demonstrably reveals the superior capabilities of the proposed framework.

\subsection{Experimental Setup}
\subsubsection {Dataset Description}

The proposed XCCNet framework was tested using four publicly available datasets, namely, Kermany Dataset~\citep{Kermany2018}, VinDr-PCXR~\citep{Pham2023vindr}, NIH-Pediatric~\citep{Wang2017nihpediatric} and Trivedi Dataset~\citep{Trivedi2022}.

\begin{figure}[!ht]
  \centering
  \includegraphics[width=0.75\linewidth]{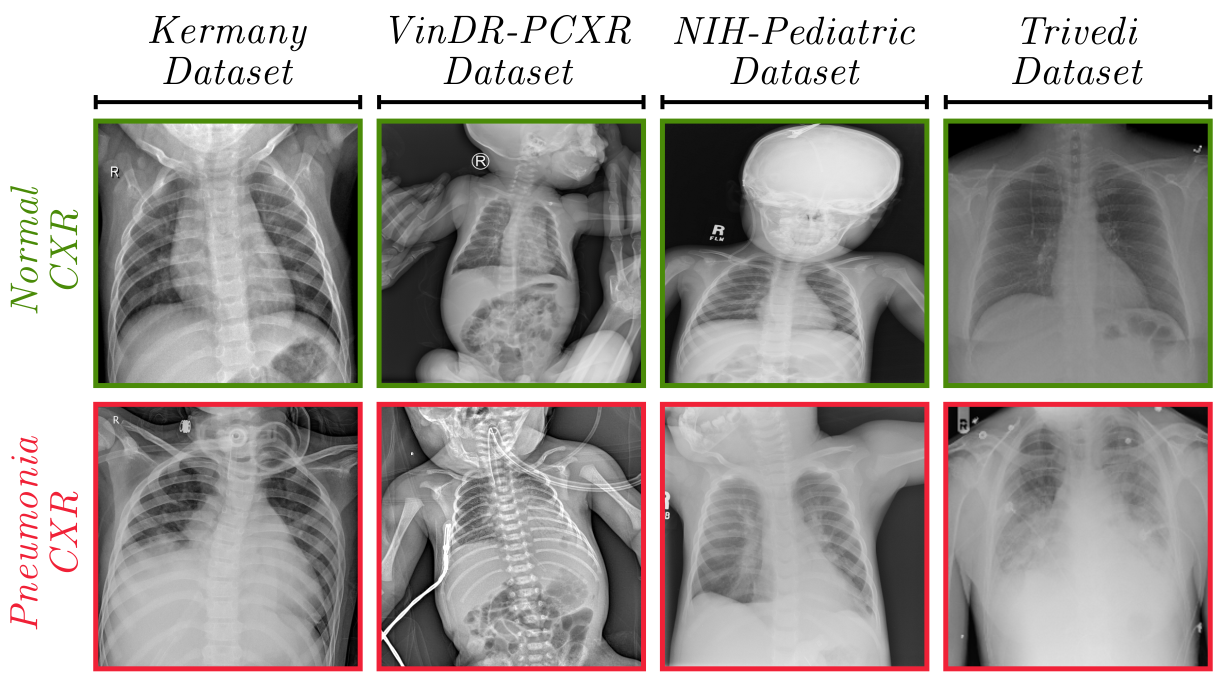}
  \caption{Sample CXRs from the Kermany, VinDR-PCXR, NIH-Pediatric, and Trivedi datasets, highlighting the diverse range of CXRs included in the study.}
  \label{fig:dataset}
\end{figure}

\autoref{fig:dataset} depicts a selection of chest radiographs utilized for effectiveness comparison of the proposed XCCNet framework. To achieve a comprehensive assessment of XCCNet generalizability, we randomly split CXR into 80\% for training, 10\% for validation, and 10\% for independent testing, as presented in~\autoref{tab:dataset}. We briefly summarize the dataset as follows:

\begin{table*}[!ht] \scriptsize
\centering
\caption{An overview of publicly available chest X-ray datasets for pediatric pneumonia detection.}
\label{tab:dataset}    
    \begin{tabular*}{0.9\linewidth}{@{\extracolsep{\fill}} lcccccc }
    \toprule
        \textbf{{Dataset}}
        &\multicolumn{2}{c}{\textbf{\thead{Original CXR$^{\#}$}}} 
        &\multicolumn{2}{c}{\textbf{\thead{Synthetic CXR$^{\ast}$}}} 
        &\multicolumn{2}{c}{\textbf{\thead{Total CXR}}} \\ 
        
        \cmidrule(l){2-3} \cmidrule(l){4-5} \cmidrule(l){6-7}
            &\textbf{Normal}&\textbf{Pneumonia}
            &\textbf{Normal}&\textbf{Pneumonia}
            &\textbf{Normal}& \textbf{Pneumonia}\\
    \midrule
        Kermany Dataset~\citep{Kermany2018} & 1583 & 4273   & 2624 & 0    & 4207 & 4273  \\
        VinDr-PCXR Dataset~\citep{Pham2023vindr} & 6050 & 1110   & 0 & 4860    & 6050 & 5970  \\
        NIH-Pediatric Dataset~\citep{Wang2017nihpediatric} & 278 & 147   & 0 & 124    & 278 & 271  \\
        Trivedi Dataset~\citep{Trivedi2022} & 1750 & 4400   & 2610 & 0    & 4360 & 4400  \\    
    \bottomrule
    \multicolumn{7}{@{}l}{$^\#$\scriptsize CXR from the original dataset $^{\ast}$\scriptsize Synthetic CXR synthesized using Adversarial-based Data Augmentation Unit}
    \end{tabular*}
\end{table*}

\paragraph[]{Kermany Dataset: \eatpunct}
The dataset provided by Kermany~\citep{Kermany2018} serves as the primary source for this study, comprising 5,856 pediatric radiographs. The dataset comprises 1,583 radiographs with ``normal'' labels and 4,273 with ``pneumonia'' labels. These chest X-ray images originate from routine screenings of pediatric patients aged 1-5 years at the Guangzhou Women and Children's Medical Centre. The dataset details are explored in~\autoref{tab:dataset}.

\paragraph[]{VinDr-PCXR Dataset: \eatpunct}
Pham \textit{et al.}~\citep{Pham2023vindr} introduce a new pediatric dataset, VinDr-PCXR, compiled from a major Vietnamese hospital between 2020 and 2021. This dataset encompasses 9,125 studies, each meticulously annotated by a pediatric radiologist with over ten years of experience. Annotations encompass 36 critical findings and 15 diseases. For our experiments, we extracted CXR labeled as ``normal'' and ``pneumonia.'' Details of the utilized dataset are presented in~\autoref{tab:dataset}.

\paragraph[]{NIH-Pediatric Dataset: \eatpunct}
Wang \textit{et al.}~\citep{Wang2017nihpediatric} present a chest X-ray dataset containing 112,120 frontal-view images of 30,805 patients. Each image has text-mined labels for 14 diseases (allowing for multi-labeling), and demographic information like age and ethnicity is also provided. The National Institutes of Health (NIH) Health Center curated the dataset and was used in this study. We specifically selected 425 images from children aged 1-5 years (matching the age range in the Guangzhou dataset). Among these, 278 were labeled as ``pneumonia'', and 147 showed ``normal'' chest findings. We refer to this subset as the "NIH-Pediatric Dataset" within our work.

\paragraph[]{Trivedi Dataset: \eatpunct}

To demonstrate the robustness of our proposed framework, we employed an external dataset utilized by Trivedi \textit{et al.}~\citep{Trivedi2022}. Due to the limited number of publicly available pediatric pneumonia datasets, we opted for this adult pneumonia dataset to assess our framework's efficacy and generalizability. It is important to note that Trivedi \textit{et al.}~\citep{Trivedi2022} employed two datasets: the Kermany dataset and an external dataset comprising pneumonia and normal chest X-rays. We specifically selected this second, unnamed external dataset for our evaluation, henceforth referred to as the ``Trivedi Dataset'' in our work. This dataset comprises 6,150 chest X-rays, of which 1,750 are labeled ``normal'' and 4,400 are labeled ``pneumonia.''

\subsubsection{Comparision Methods}
\label{sec:comparisionmethods}
In this section, we offer a concise overview of the existing literature, focusing on the key techniques and functionalities employed by these methods. This overview will serve as a benchmark for evaluating the performance of the XCCNet framework.

\paragraph[]{i) Convolutional-based Pneumonia Detection:\eatpunct}

We examine the performance of existing convolutional-based methods as baseline methods for the detection of pediatric pneumonia. Trivedi \textit{et al.}~\citep{Trivedi2022} propose a lightweight Deep Learning architecture built on MobileNet. This architecture primarily uses depth-wise separable convolutions alongside GlobalAveragePooling2D for automated pneumonia detection. Kili{\c{c}}arslan \textit{et al.}~\citep{Kilicarslan2023} develop a novel activation function, SupEx, which combines TanhExp and ReLU algebraically. This preserves the crucial differentiability property of TanhExp, influencing subsequent layers within the CNN during backpropagation. Yi \textit{et al.}~\citep{Yi2023dl} introduce a scalable and interpretable Deep Convolutional Neural Network (DCNN) specifically designed to extract primary features from CXR images for pneumonia identification.

\paragraph[]{ii) Transformer-based Pneumonia Detection:\eatpunct}

This section examines existing transformer-based baselines for pediatric pneumonia detection.  Usman \textit{et al.}~\citep{Usman2022} leverage a pre-trained vision transformer on both the CheXpert and a pediatric pneumonia dataset. By applying random weight initialization, they aim to learn long-range features from CXR images. Okolo \textit{et al.}~\citep{Okolo2022} combine features from a ResNet-based skip connection with those extracted by a vision transformer. This parallel integration allows their model to learn both local and global information within CXR images for pneumonia classification. Singh \textit{et al.}~\citep{Singh2024evit} utilizes a vision transformer for pneumonia detection in CXR images, emphasizing its efficiency in capturing global features.

\paragraph[]{iii) Hybrid Model-based Pneumonia Detection:\eatpunct}

Focusing on pediatric pneumonia detection, this work examines several baseline approaches that combine convolutional and transformer-based methods. Ayan \textit{et al.}~\citep{Ayan2022} propose an automatic pneumonia diagnosis system for children using an ensemble of convolutional neural networks (CNNs) including Xception, ResNet-50, and MobileNet. To prevent spatial feature loss, they add a global average pooling layer. Chattopadhyay \textit{et al.}~\citep{Chattopadhyay2022} introduces a feature selection framework using the Sine Cosine Algorithm (SCA) to enhance pneumonia detection. Features are extracted using a pre-trained DenseNet-201, followed by SCA for selection. Prakash \textit{et al.}~\citep{Prakash2023tl} propose stacked ensemble learning on features extracted from ResNet50V2, ResNet101V2, ResNet152V2, Xception, and DenseNet169. Additionally, Kernel PCA reduces dimensionality, and a stacking classifier with XGBClassifier, Support Vector Classifier, Logistic Regression, Nu-SVC, and K-Nearest Neighbour performs classification. They then classify using a Nu-SVC meta-classifier. Kaya \textit{et al.}~\citep{Kaya2024featurefusion} develop a hierarchical template-matching approach for relevant feature learning. Features are extracted from multiple CNN models with different architectures and classified using majority voting. Lie \textit{et al.}~\citep{Liu2024effctm} propose a hierarchical mechanism combining CNN, Transformer, and Multi-Layer Perceptron (MLP). This approach integrates spatial and global feature extraction.

\paragraph[]{iv) Explainable-based Pneumonia Detection:\eatpunct}

This section examines existing baselines leveraging explainability that offer insights into feature learnings for pediatric pneumonia detection. Ukwuoma \textit{et al.}~\citep{Ukwuoma2023xet} propose an explainable hybrid model that combines deep ensemble techniques (GoogleNet, DenseNet201, and VGG16) with a transformer encoder for accurate CXR-based pneumonia detection. This integration provides explainable maps, aiding in understanding the model's predictions. Yang \textit{et al.}~\citep{Yang2022xai} present an explainable deep learning approach that removes lung backgrounds. They utilize GradCAM to generate explainability maps from a fine-tuned VGG16 model, allowing users to understand which regions contribute most to the prediction. Hroub \textit{et al.}~\citep{Hroub2024xdl} demonstrate the effectiveness of efficient data augmentation with InceptionV3, surpassing vision transformers in performance. Additionally, they visually illustrate their technique's decision-making process using AblationCAM, EigenGradCAM, GradCam++, RandomCAM, and GradCAM.

\subsubsection{Evaluation Metrices}

The XCCNet framework, designed to detect pneumonia in chest radiographs and categorize each image as either pneumonia or normal, requires thorough evaluation to assess its feasibility. We employ four key metrics: accuracy, precision, recall, and F1-score. As our dataset avoids class imbalance, focusing on accuracy and F1-score provides a robust assessment of XCCNet's overall performance. Additionally, we consider precision and recall to gain insights into the model's ability to identify pneumonia cases and avoid misclassifying normal individuals accurately. \autoref{tab:evalmetrices} shows the mathematical representations of the metrics.

\begin{table}[!ht] \scriptsize
\centering
\caption{Evaluation metrics for XCCNet framework in the detection of pediatric pneumonia from CXR.}
\label{tab:evalmetrices}    
    \begin{tabular*}{0.7\linewidth}{@{\extracolsep{\fill}} lc}
    \toprule
        \textbf{Evaluation Metrics} & \textbf{Mathematical Formulation} \\
        \midrule
        Accuracy & $ \mathlarger{\frac{T\!P_{pneumonia}+T\!N_{normal}}{T\!P_{pneumonia}+F\!P_{pneumonia}+T\!N_{normal}+F\!N_{normal}}}$  \\
        & \\
        Precision & $ \mathlarger{\frac{T\!P_{pneumonia}}{T\!P_{pneumonia} +F\!P_{pneumonia}}}$  \\
        & \\
        Recall & $ \mathlarger{\frac{T\!P_{pneumonia}}{T\!P_{pneumonia}+F\!N_{normal}}}$  \\
        & \\
        F1 - Score & $ \mathlarger{ 2 \times \frac{ Precision \times Recall}{Precision + Recall}}$ \\
    \bottomrule
    \end{tabular*}
\end{table}

Here, $T\!P_{penumonia}$ is True Positive, a pneumonia CXR correctly predicted as pneumonia. $T\!N_{normal}$ is True Negative, a normal CXR correctly predicted as normal. $F\!P_{penumonia}$ is False Positive, a normal CXR incorrectly predicted as pneumonia. $F\!N_{normal}$ is False Negative, a pneumonia CXR incorrectly predicted as normal.

\subsubsection{Implementation Details}

Our experiments were conducted on an NVIDIA P100 GPU with 16GB GPU RAM and 512GB system RAM using Python and the TensorFlow library. To optimize model training, we employed the Adam optimizer with an initial learning rate of 1e-6 and implemented an adaptive learning rate strategy. This strategy utilized a callback function to monitor validation accuracy and automatically reduced the learning rate by 0.3 upon stagnation. Adversarial-based data augmentation was incorporated to counter overfitting. This involved generating high-quality synthetic CXRs from real samples, thereby diversifying the training set and enhancing model generalizability to unseen data. This augmentation improved the model's robustness and minimized the risk of overfitting to specific training patterns. To further enhance stability and promote sparsity, both L1 and L2 regularization were utilized during training. L1 regularization enforced a Laplace before the weights, encouraging the model to rely on a select subset of influential features, while L2 regularization constrained weight magnitudes to prevent overfitting.

\subsection{Results and Analysis}

This section presents a comprehensive comparison of the proposed XCCNet framework against state-of-the-art methods. Next, we perform targeted ablation studies, systematically isolating and evaluating the individual contributions of each module within the framework. This analysis pinpoints the impact and efficacy of each component on the overall performance. Finally, we explore explainability analysis to visually illustrate how XCCNet identifies primary areas and learns features within chest X-rays.

\subsubsection{Comparison with State-of-the-Art Methods}

In this section, we assess the performance of our proposed Explainable Contrastive-based Dilated Convolutional Network with Transformer (XCCNet) against fourteen state-of-the-art (SOTA) (detailed in~\autoref{sec:comparisionmethods}). We leverage four diverse datasets to ensure generalizability and employ accuracy, precision, recall, and F1-score as evaluation metrics since our datasets exhibit balanced class distributions. Higher values in these metrics indicate superior classification performance. For a comprehensive comparison, we categorize existing state-of-the-art methods into four groups: Convolutional-based, Transformer-based, Hybrid Model-based, and Explainable-based approaches. Notably, XCCNet outperforms all competitors in terms of accuracy on the Kermany, NIH-Pediatric, and Trivedi datasets. Furthermore, on the VinDr-PCXR dataset, XCCNet achieves the second-highest accuracy, showcasing its competitive performance across various data sources.

\paragraph[]{Performance Evaluation on Kermany Dataset:\eatpunct}

This section evaluates the proposed XCCNet against established baseline methods on the Kermany dataset, with results presented in~\autoref{tab:kermany}. Our XCCNet achieves outstanding performance, demonstrating an accuracy of 99.76\%, precision of 99.75\%, recall of 99.75\%, and F1-score of 99.75\%. XCCNet significantly surpasses the second-best method by Ukwuoma \textit{et al.}~\cite{Ukwuoma2023xet} in terms of accuracy, recall, and F1-score, with margins of 0.55\%, 0.54\%, and 0.54\%, respectively. While comparable precision is observed with Okolo \textit{et al.}~\cite{Okolo2022}, it demonstrates the effectiveness of transformer-based methods for achieving high precision. While many existing baselines focused their evaluation on the Kermany dataset, we conducted a broader evaluation by reproducing their results from scratch and applying them to additional diverse datasets. This comprehensive comparison across different data distributions demonstrates the true generalizability and effectiveness of the XCCNet framework compared to existing methods. Notably, explainability-based techniques also exhibit promising performance. However, XCCNet's unique integration of transformers and convolution-based extraction enables it to capture both fine-grained details and long-range features for superior feature extraction, highlighting its effectiveness compared to existing methods. 

\begin{table*}[!ht] \scriptsize
\centering
\caption{Performance comparison of XCCNet with state-of-the-art methods for pediatric pneumonia detection on the Kermany dataset. Higher values (accuracy, precision, recall, F1-Score) signify superior performance. The best and second-best performing methods are highlighted in \colorbox{inchworm}{green} and \colorbox{yellow}{yellow}. $|\nabla|$ denotes the absolute performance drop relative to XCCNet.}
% \caption{Comparison of XCCNet and state-of-the-art methods for pediatric pneumonia detection on the Kermany dataset. Higher values of accuracy, precision, recall, and F1-Score indicate better performance. Best and second-best results are highlighted in \colorbox{inchworm}{green} and \colorbox{yellow}{yellow}, respectively. $|\nabla|$ denotes the absolute performance drop relative to XCCNet.}
    % \begin{tabular}{l cc cc cc cc}        
    \begin{tabular*}{0.9\linewidth}{@{\extracolsep{\fill}}l cc cc cc cc}
    \toprule
    \textbf{Methods} & \textbf{Acc$^a$} (\%) & $|\nabla|$ & \textbf{ Pre$^b$} (\%) & $|\nabla|$ & \textbf{ Rec$^c$} (\%) & $|\nabla|$ & \textbf{ F1$^d$} (\%) & $|\nabla|$\\                  
    \midrule
    \multicolumn{9}{l}{\textbf{Convolutional-based Pneumonia Detection}}\\        
    \midrule        
    Trivedi \textit{et al.}~\cite{Trivedi2022}&94.23 & 5.53 & 95.00 & 4.75 & 96.00 & 3.75 & 95.00 & 4.75 \\
    Kili{\c{c}}arslan \textit{et al.}~\cite{Kilicarslan2023}&95.37 & 4.39 & - & - & - & - & - & - \\
    Yi \textit{et al.}~\cite{Yi2023dl}&96.06 & 3.70 & - & - & 93.58 & 6.17 & - & - \\
        
    \midrule
    \multicolumn{9}{l}{\textbf{Transformer-based Pneumonia Detection}} \\
    \midrule
    Usman \textit{et al.}~\cite{Usman2022}&88.46 & 11.30 & 83.75 & 16.00 & 85.90 & 13.85 & 84.81 & 14.94 \\
    Okolo \textit{et al.}~\cite{Okolo2022}&98.08 & 1.68 & \colorbox{yellow}{99.74} & 0.01 & 97.25 & 2.50 & 98.48 & 1.27 \\
    Singh \textit{et al.}~\cite{Singh2024evit}&97.61 & 2.15 & - & - & - & - & 95.00 & 4.75 \\

    \midrule
    \multicolumn{9}{l}{\textbf{Hybrid Model-based Pneumonia Detection}} \\
    \midrule
    Ayan \textit{et al.}~\cite{Ayan2022}&95.83 & 3.93 & 96.02 & 3.73 & 92.74 & 7.01 & 94.35 & 5.40 \\
    Chattopadhyay \textit{et al.}~\cite{Chattopadhyay2022}&98.36 & 1.40 & 98.98 & 0.77 & 98.79 & 0.96 & 98.80 & 0.95 \\
    Prakash \textit{et al.}~\cite{Prakash2023tl}&96.15 & 3.61 & 97.91 & 1.84 & 95.90 & 3.85 & 96.24 & 3.51 \\
    Kaya \textit{et al.}~\cite{Kaya2024featurefusion}&98.94 & 0.82 & 99.12 & 0.63 & 99.12 & 0.63 & 99.12 & 0.63 \\
    Liu \textit{et al.}~\cite{Liu2024effctm}&97.03 & 2.73 & - & - & - & - & 97.63 & 2.12 \\
        
    \midrule
    \multicolumn{9}{l}{\textbf{Explainable-based Pneumonia Detection}} \\
    \midrule
    Ukwuoma \textit{et al.}~\cite{Ukwuoma2023xet}& \colorbox{yellow}{99.21} & 0.55 & 99.21 & 0.54 & \colorbox{yellow}{99.21} & 0.54 & \colorbox{yellow}{99.21} & 0.54 \\
    Yang \textit{et al.}~\cite{Yang2022xai}&95.60 & 4.16 & - & - & - & - & - & - \\
    Hroub \textit{et al.}~\cite{Hroub2024xdl}&95.55 & 4.21 & 94.54 & 5.21 & 94.55 & 5.20 & 94.54 & 5.21 \\

    \midrule
    \multicolumn{9}{l}{\textbf{Explainable Contrastive-based Dilated Convolutional Network with Transformer (Ours)}} \\
    \midrule
    XCCNet&\colorbox{inchworm}{99.76} & 0.00 & \colorbox{inchworm}{99.75} & 0.00 & \colorbox{inchworm}{99.75} & 0.00 & \colorbox{inchworm}{99.75} & 0.00\\
    \bottomrule
    \multicolumn{9}{l}{\scriptsize $^a$ Accuracy, $^b$ Precision, $^c$ Recall, $^d$ F1-Score, `-' no result reported}
    % \end{tabular}
    \end{tabular*}
    \label{tab:kermany}
\end{table*}

\paragraph[]{Performance Evaluation on VinDR-CXR Dataset:\eatpunct}

\autoref{tab:vindrpcxr} showcases the comparison results on the VinDr-PCXR dataset. While the method proposed by Okolo \textit{et al.}~\cite{Okolo2022} achieves the highest accuracy (0.01\% margin), it shares similarities with XCCNet, including the integration of vision transformers and CNNs and the use of skip-connections for feature fusion. Notably, XCCNet surpasses Ayan \textit{et al.}~\cite{Ayan2022}, Kaya \textit{et al.}~\cite{Kaya2024featurefusion}, and even Okolo \textit{et al.}~\cite{Okolo2022} in terms of precision (by 0.30\%), recall (by 0.06\%), and F1-Score (by 0.64\%). This demonstrates XCCNet's effectiveness even against strong contenders. While Ayan \textit{et al.}~\cite{Ayan2022} and Kaya \textit{et al.}~\cite{Kaya2024featurefusion} leverage hybrid approaches for pneumonia feature learning, XCCNet's integrated adversarial-based data augmentation unit effectively tackles data skewness. This capability leads to more accurate identification of pneumonia cases and minimizes the risk of misclassifying normal CXR images.

\begin{table*}[!ht] \scriptsize
\centering
\caption{Performance comparison of XCCNet with state-of-the-art methods for pediatric pneumonia detection on the VinDR-PCXR dataset. Higher values (accuracy, precision, recall, F1-Score) signify superior performance. The best and second-best performing methods are highlighted in \colorbox{inchworm}{green} and \colorbox{yellow}{yellow}. $|\nabla|$ denotes the absolute performance drop relative to XCCNet.}
% \caption{Comparison of XCCNet and state-of-the-art methods for pediatric pneumonia detection on the VinDr-PCXR dataset. Higher values of accuracy, precision, recall, and F1-Score indicate better performance. Best and second-best results are highlighted in \colorbox{inchworm}{green} and \colorbox{yellow}{yellow}, respectively. $|\nabla|$ denotes the absolute performance drop relative to XCCNet.}
    \begin{tabular*}{0.9\linewidth}{@{\extracolsep{\fill}}l cc cc cc cc}
    \toprule
    \textbf{Methods} & \textbf{Acc$^a$} (\%) & $|\nabla|$ & \textbf{ Pre$^b$} (\%) & $|\nabla|$ & \textbf{ Rec$^c$} (\%) & $|\nabla|$ & \textbf{ F1$^d$} (\%) & $|\nabla|$\\                 
    \midrule
    \multicolumn{9}{l}{\textbf{Convolutional-based Pneumonia Detection}}\\        
    \midrule        
    Trivedi \textit{et al.}~\cite{Trivedi2022}&87.28 & 4.28 & 91.83 & 0.35 & 84.56 & 5.41 & 88.05 & 3.01 \\
    Kili{\c{c}}arslan \textit{et al.}~\cite{Kilicarslan2023}&89.43 & 2.13 & 90.78 & 1.40 & 86.37 & 3.60 & 88.52 & 2.54 \\
    Yi \textit{et al.}~\cite{Yi2023dl}&89.72 & 1.84 & 90.21 & 1.97 & 87.57 & 2.40 & 88.87 & 2.19 \\
        
    \midrule
    \multicolumn{9}{l}{\textbf{Transformer-based Pneumonia Detection}} \\
    \midrule
    Usman \textit{et al.}~\cite{Usman2022}&86.34 & 5.22 & 90.79 & 1.39 & 85.81 & 4.16 & 88.23 & 2.83 \\
    Okolo \textit{et al.}~\cite{Okolo2022}&\colorbox{inchworm}{91.57} & -0.01 & 91.38 & 0.80 & 89.48 & 0.49 & \colorbox{yellow}{90.42} & 0.64 \\
    Singh \textit{et al.}~\cite{Singh2024evit}&89.57 & 1.99 & 89.40 & 2.78 & 86.54 & 3.43 & 87.94 & 3.12 \\

    \midrule
    \multicolumn{9}{l}{\textbf{Hybrid Model-based Pneumonia Detection}} \\
    \midrule
    Ayan \textit{et al.}~\cite{Ayan2022}&86.72 & 4.84 & \colorbox{yellow}{91.88} & 0.30 & 84.79 & 5.18 & 88.19 & 2.87 \\
    Chattopadhyay \textit{et al.}~\cite{Chattopadhyay2022}&91.52 & 0.04 & 89.73 & 2.45 & 88.96 & 1.01 & 89.34 & 1.72 \\
    Prakash \textit{et al.}~\cite{Prakash2023tl}&87.45 & 4.11 & 91.05 & 1.13 & 85.34 & 4.63 & 88.10 & 2.96 \\
    Kaya \textit{et al.}~\cite{Kaya2024featurefusion}&91.38 & 0.18 & 90.77 & 1.41 & \colorbox{yellow}{89.91} & 0.06 & 90.34 & 0.72 \\
    Liu \textit{et al.}~\cite{Liu2024effctm}&89.18 & 2.38 & 91.03 & 1.15 & 87.52 & 2.45 & 89.24 & 1.82 \\
        
    \midrule
    \multicolumn{9}{l}{\textbf{Explainable-based Pneumonia Detection}} \\
    \midrule
    Ukwuoma \textit{et al.}~\cite{Ukwuoma2023xet}&91.27 & 0.29 & 90.61 & 1.57 & 89.19 & 0.78 & 89.89 & 1.17 \\
    Yang \textit{et al.}~\cite{Yang2022xai}&87.39 & 4.17 & 90.95 & 1.23 & 85.81 & 4.16 & 88.30 & 2.76 \\
    Hroub \textit{et al.}~\cite{Hroub2024xdl}&88.54 & 3.02 & 91.28 & 0.90 & 83.25 & 6.72 & 87.08 & 3.98 \\

    \midrule
    \multicolumn{9}{l}{\textbf{Explainable Contrastive-based Dilated Convolutional Network with Transformer (Ours)}} \\
    \midrule
    XCCNet&\colorbox{yellow}{91.56} & 0.00 & \colorbox{inchworm}{92.18} & 0.00 & \colorbox{inchworm}{89.97} & 0.00 & \colorbox{inchworm}{91.06} & 0.00\\
    \bottomrule
    \multicolumn{9}{l}{\scriptsize $^a$ Accuracy, $^b$ Precision, $^c$ Recall, $^d$ F1-Score}
    % \end{tabular}
    \end{tabular*}
    \label{tab:vindrpcxr}
\end{table*}

\paragraph[]{Performance Evaluation on NIH-Pediatric Dataset:\eatpunct}

As shown in~\autoref{tab:nihped}, the proposed XCCNet framework surpasses all other state-of-the-art (SOTA) techniques on the NIH-Pediatric dataset in terms of accuracy, precision, and F1-score. It achieves impressive values of 92.87\%, 91.72\%, and 91.65\%, respectively. While the hybrid feature extraction technique by Kaya \textit{et al.}~\cite{Kaya2024featurefusion} exhibits comparable performance, securing the second-best results across all metrics, its approach relies on an ensemble strategy with majority voting. This contrasts with XCCNet's robust preprocessing module, which effectively amplifies low-radiation CXRs into high-quality images. This enhanced input allows XCCNet to learn deeper features and extract valuable insights, ultimately leading to superior performance.

\begin{table*}[!ht] \scriptsize
\centering
\caption{Performance comparison of XCCNet with state-of-the-art methods for pediatric pneumonia detection on the NIH-Pediatric dataset. Higher values (accuracy, precision, recall, F1-Score) signify superior performance. The best and second-best performing methods are highlighted in \colorbox{inchworm}{green} and \colorbox{yellow}{yellow}. $|\nabla|$ denotes the absolute performance drop relative to XCCNet.}
% \caption{Comparison of XCCNet and state-of-the-art methods for pediatric pneumonia detection on the NIH-Pediatric dataset. Higher values of accuracy, precision, recall, and F1-Score indicate better performance. Best and second-best results are highlighted in \colorbox{inchworm}{green} and \colorbox{yellow}{yellow}, respectively. $|\nabla|$ denotes the absolute performance drop relative to XCCNet.}
    \begin{tabular*}{0.9\linewidth}{@{\extracolsep{\fill}}l cc cc cc cc}
    \toprule
    \textbf{Methods} & \textbf{Acc$^a$} (\%) & $|\nabla|$ & \textbf{ Pre$^b$} (\%) & $|\nabla|$ & \textbf{ Rec$^c$} (\%) & $|\nabla|$ & \textbf{ F1$^d$} (\%) & $|\nabla|$\\                  
    \midrule
    \multicolumn{9}{l}{\textbf{Convolutional-based Pneumonia Detection}}\\        
    \midrule        
    Trivedi \textit{et al.}~\cite{Trivedi2022}&85.45 & 7.42 & 86.66 & 5.06 & 87.78 & 3.81 & 87.22 & 4.43 \\
    Kili{\c{c}}arslan \textit{et al.}~\cite{Kilicarslan2023}&87.71 & 5.16 & 87.23 & 4.49 & 89.14 & 2.45 & 88.17 & 3.48 \\
    Yi \textit{et al.}~\cite{Yi2023dl}&84.22 & 8.65 & 86.33 & 5.39 & 87.18 & 4.41 & 86.76 & 4.89 \\
        
    \midrule
    \multicolumn{9}{l}{\textbf{Transformer-based Pneumonia Detection}} \\
    \midrule
    Usman \textit{et al.}~\cite{Usman2022}&86.67 & 6.2 & 86.98 & 4.74 & 88.91 & 2.68 & 87.93 & 3.72 \\
    Okolo \textit{et al.}~\cite{Okolo2022}&89.54 & 3.33 & 88.91 & 2.81 & 90.26 & 1.33 & 89.58 & 2.07 \\
    Singh \textit{et al.}~\cite{Singh2024evit}&88.19 & 4.68 & 87.66 & 4.06 & 89.27 & 2.32 & 88.46 & 3.19 \\
    \midrule
    \multicolumn{9}{l}{\textbf{Hybrid Model-based Pneumonia Detection}} \\
    \midrule
    Ayan \textit{et al.}~\cite{Ayan2022}&91.08 & 1.79 & 89.67 & 2.05 & 91.23 & 0.36 & 90.44 & 1.21 \\
    Chattopadhyay \textit{et al.}~\cite{Chattopadhyay2022}&91.34 & 1.53 & 88.99 & 2.73 & 90.82 & 0.77 & 89.89 & 1.76 \\
    Prakash \textit{et al.}~\cite{Prakash2023tl}&88.48 & 4.39 & 88.14 & 3.58 & 89.37 & 2.22 & 88.75 & 2.90 \\
    Kaya \textit{et al.}~\cite{Kaya2024featurefusion}&\colorbox{yellow}{92.68} & 0.19 & \colorbox{yellow}{90.80} & 0.92 & \colorbox{inchworm}{92.07} & 0.48 & \colorbox{yellow}{91.43} & 0.22 \\
    Liu \textit{et al.}~\cite{Liu2024effctm}&91.09 & 1.78 & 88.93 & 2.79 & 91.43 & 0.16 & 90.16 & 1.49 \\
        
    \midrule
    \multicolumn{9}{l}{\textbf{Explainable-based Pneumonia Detection}} \\
    \midrule
    Ukwuoma \textit{et al.}~\cite{Ukwuoma2023xet}&91.71 & 1.16 & 90.20 & 1.52 & 91.58 & 0.01 & 90.89 & 0.76 \\
    Yang \textit{et al.}~\cite{Yang2022xai}&88.36 & 4.51 & 88.10 & 3.62 & 89.00 & 2.59 & 88.55 & 3.10 \\
    Hroub \textit{et al.}~\cite{Hroub2024xdl}&87.05 & 5.82 & 87.16 & 4.56 & 89.28 & 2.31 & 88.20 & 3.45 \\

    \midrule
    \multicolumn{9}{l}{\textbf{Explainable Contrastive-based Dilated Convolutional Network with Transformer (Ours)}} \\
    \midrule
    XCCNet&\colorbox{inchworm}{92.87} & 0.00 & \colorbox{inchworm}{91.72} & 0.00 & \colorbox{yellow}{91.59} & 0.00 & \colorbox{inchworm}{91.65} & 0.00\\ 
    \bottomrule
    \multicolumn{9}{l}{\scriptsize $^a$ Accuracy, $^b$ Precision, $^c$ Recall, $^d$ F1-Score}
    % \end{tabular}
    \end{tabular*}
    \label{tab:nihped}
\end{table*}

\paragraph[]{Performance Evaluation on Trivedi Dataset:\eatpunct}

This section presents the proposed XCCNet's performance on the Trivedi dataset against established baselines (refer~\autoref{tab:trivedi}). XCCNet achieves the highest accuracy, precision, and F1-score, demonstrating exceptional results with values of 97.19\%, 96.72\%, and 97.00\%, respectively. While Okolo \textit{et al.}~\cite{Okolo2022} edge out a slightly higher recall of 97.85\%, their approach heavily relies on integrating convolutions with transformers, which may not generalize as effectively across diverse datasets. XCCNet surpasses competitors in other metrics: Liu \textit{et al.}~\cite{Liu2024effctm} for accuracy and F1-score, and Ukwuoma \textit{et al.}~\cite{Ukwuoma2023xet} for precision. Notably, the margins between the best and second-best results are minimal (0.83\% for accuracy, 0.17\% for precision, and 0.69\% for F1-Score), showcasing the competitive performance. Compared to existing techniques, XCCNet utilizes contrastive learning with binary cross-entropy. This approach effectively enhances feature learning by considering both inter-class similarities and dissimilarities, ultimately leading to superior classification performance and accurate differentiation between pneumonia and normal cases.

\begin{table*}[!ht] \scriptsize
\centering
\caption{Performance comparison of XCCNet with state-of-the-art methods for pediatric pneumonia detection on the Trivedi dataset. Higher values (accuracy, precision, recall, F1-Score) signify superior performance. The best and second-best performing methods are highlighted in \colorbox{inchworm}{green} and \colorbox{yellow}{yellow}. $|\nabla|$ denotes the absolute performance drop relative to XCCNet.}
% \caption{Comparison of XCCNet and state-of-the-art methods for pediatric pneumonia detection on the Trivedi dataset. Higher values of accuracy, precision, recall, and F1-Score indicate better performance. Best and second-best results are highlighted in \colorbox{inchworm}{green} and \colorbox{yellow}{yellow}, respectively. $|\nabla|$ denotes the absolute performance drop relative to XCCNet.}
    \begin{tabular*}{0.9\linewidth}{@{\extracolsep{\fill}}l cc cc cc cc}
    \toprule
    \textbf{Methods} & \textbf{Acc$^a$} (\%) & $|\nabla|$ & \textbf{ Pre$^b$} (\%) & $|\nabla|$ & \textbf{ Rec$^c$} (\%) & $|\nabla|$ & \textbf{ F1$^d$} (\%) & $|\nabla|$\\                  
    \midrule
    \multicolumn{9}{l}{\textbf{Convolutional-based Pneumonia Detection}}\\        
    \midrule        
    Trivedi \textit{et al.}~\cite{Trivedi2022}&87.46 & 9.73 & 90.64 & 6.08 & 89.71 & 7.57 & 90.17 & 6.83 \\
    Kili{\c{c}}arslan \textit{et al.}~\cite{Kilicarslan2023}&94.43 & 2.76 & 94.78 & 1.94 & 96.68 & 0.60 & 95.72 & 1.28 \\
    Yi \textit{et al.}~\cite{Yi2023dl}&90.46 & 6.73 & 94.79 & 1.93 & 95.48 & 1.80 & 95.13 & 1.87 \\
        
    \midrule
    \multicolumn{9}{l}{\textbf{Transformer-based Pneumonia Detection}} \\
    \midrule
    Usman \textit{et al.}~\cite{Usman2022}&94.43 & 2.76 & 93.13 & 3.59 & 96.70 & 0.58 & 94.88 & 2.12 \\
    Okolo \textit{et al.}~\cite{Okolo2022}&95.86 & 1.33 & 94.42 & 2.30 & \colorbox{inchworm}{97.85} & 0.57 & 96.10 & 0.90 \\
    Singh \textit{et al.}~\cite{Singh2024evit}&88.52 & 8.67 & 92.91 & 3.81 & 92.41 & 4.87 & 92.66 & 4.34 \\

    \midrule
    \multicolumn{9}{l}{\textbf{Hybrid Model-based Pneumonia Detection}} \\
    \midrule
    Ayan \textit{et al.}~\cite{Ayan2022}&95.73 & 1.46 & 92.41 & 4.31 & 96.05 & 1.23 & 94.19 & 2.81 \\
    Chattopadhyay \textit{et al.}~\cite{Chattopadhyay2022}&96.33 & 0.86 & 95.22 & 1.50 & 95.52 & 1.76 & 95.37 & 1.63 \\
    Prakash \textit{et al.}~\cite{Prakash2023tl}&94.77 & 2.42 & 92.49 & 4.23 & 93.53 & 3.75 & 93.01 & 3.99 \\
    Kaya \textit{et al.}~\cite{Kaya2024featurefusion}&96.08 & 1.11 & 95.08 & 1.64 & 96.56 & 0.72 & 95.81 & 1.19 \\
    Liu \textit{et al.}~\cite{Liu2024effctm}&\colorbox{yellow}{96.36} & 0.83 & 96.22 & 0.50 & 96.41 & 0.87 & \colorbox{yellow}{96.31} & 0.69 \\
        
    \midrule
    \multicolumn{9}{l}{\textbf{Explainable-based Pneumonia Detection}} \\
    \midrule
    Ukwuoma \textit{et al.}~\cite{Ukwuoma2023xet}&95.61 & 1.58 & \colorbox{yellow}{96.55} & 0.17 & 94.36 & 2.92 & 95.44 & 1.56 \\
    Yang \textit{et al.}~\cite{Yang2022xai}&92.49 & 4.70 & 94.91 & 1.81 & 91.54 & 5.74 & 93.19 & 3.81 \\
    Hroub \textit{et al.}~\cite{Hroub2024xdl}&93.14 & 4.05 & 93.71 & 3.01 & 95.67 & 1.61 & 94.68 & 2.32 \\

    \midrule
    \multicolumn{9}{l}{\textbf{Explainable Contrastive-based Dilated Convolutional Network with Transformer (Ours)}} \\
    \midrule
    XCCNet&\colorbox{inchworm}{97.19} & 0.00 & \colorbox{inchworm}{96.72} & 0.00 & \colorbox{yellow}{97.28} & 0.00 & \colorbox{inchworm}{97.00} & 0.00\\
    \bottomrule
    \multicolumn{9}{l}{\scriptsize $^a$ Accuracy, $^b$ Precision, $^c$ Recall, $^d$ F1-Score}
    % \end{tabular}
    \end{tabular*}
    \label{tab:trivedi}
\end{table*}

\paragraph[]{Comparision with Existing Transfer Learning Techniques:\eatpunct}

\begin{table*}[!ht] \scriptsize
\centering
\caption{Performance comparison of XCCNet with existing transfer learning techniques for pediatric pneumonia detection on the Kermany dataset. Higher values (accuracy, precision, recall, F1-Score) signify superior performance. The best and second-best performing methods are highlighted in \colorbox{inchworm}{green} and \colorbox{yellow}{yellow}. $|\nabla|$ denotes the absolute performance drop relative to XCCNet.}
% \caption{Comparison of XCCNet with existing transfer learning techniques for pediatric pneumonia detection on the Kermany dataset. Higher values of accuracy, precision, recall, and F1-Score indicate better performance. Best and second-best results are highlighted in \colorbox{inchworm}{green} and \colorbox{yellow}{yellow}, respectively. $|\nabla|$ denotes the absolute performance drop relative to XCCNet.}
    \begin{tabular*}{0.9\linewidth}{@{\extracolsep{\fill}}l cc cc cc cc}
    \toprule
    \textbf{Methods} & \textbf{Acc$^a$} (\%) & $|\nabla|$ & \textbf{ Pre$^b$} (\%) & $|\nabla|$ & \textbf{ Rec$^c$} (\%) & $|\nabla|$ & \textbf{ F1$^d$} (\%) & $|\nabla|$\\                  
    \midrule
    % \multicolumn{9}{l}{\textbf{Convolutional-based Pneumonia Detection}}\\        
    % \midrule        
    DenseNet121 & 95.27 & 4.49 & 92.43 & 7.32 & 98.24 & 1.51 & 95.25 & 4.50 \\
    DenseNet169 & 92.97 & 6.79 & 87.67 & 12.08 & \colorbox{yellow}{99.50} & 0.25 & 93.18 & 6.57 \\
    VGG16 & 97.58 & 2.18 & 95.87 & 3.88 & 99.25 & 0.50 & 97.53 & 2.22 \\
    Xception & 97.21 & 2.55 & 95.62 & 4.13 & 98.74 & 1.01 & 97.16 & 2.59 \\
    ResNet50\_v2 & 89.33 & 10.43 & 86.38 & 13.37 & 92.46 & 7.29 & 89.32 & 10.43 \\
    ResNet152\_v2 & \colorbox{yellow}{98.18} & 1.58 & \colorbox{yellow}{96.59} & 3.16 & \colorbox{inchworm}{99.75} & 0.00 & \colorbox{yellow}{98.15} & 1.60 \\
    EfficientNet\_B0\_vl & 95.03 & 4.73 & 94.07 & 5.68 & 95.73 & 4.02 & 94.89 & 4.86 \\
    EfficientNet\_B0\_v2 & 97.94 & 1.82 & 94.04 & 5.71 & 98.74 & 1.01 & 97.88 & 1.87 \\
    \midrule
    XCCNet & \colorbox{inchworm}{99.76} & 0.00 & \colorbox{inchworm}{99.75} & 0.00 & \colorbox{inchworm}{99.75} & 0.00 & \colorbox{inchworm}{99.75} & 0.00\\
    \bottomrule
    \multicolumn{9}{l}{\scriptsize $^a$ Accuracy, $^b$ Precision, $^c$ Recall, $^d$ F1-Score}
    % \end{tabular}
    \end{tabular*}
    \label{tab:transferlearning}
\end{table*}

To assess the effectiveness of XCCNet, we evaluated the performance of eight popular transfer learning techniques on the Kermany dataset. These techniques included Densenet121, DenseNet169, VGG16, Xception, ResNet50\_v2, ResNet152\_v2, EfficientNet\_B0\_v1, and EfficientNet\_B0\_v2. All models were fine-tuned on the Kermany dataset for optimal performance. As shown in~\autoref{tab:transferlearning}, XCCNet outperforms all compared pre-trained models. While ResNet152\_v2 achieves the closest performance, reaching an accuracy of 98.18\%, precision of 96.59\%, and F1-score of 98.15\%, it falls short of XCCNet's accuracy by 1.58\% and F1-score by 1.6\%. Notably, both XCCNet and ResNet152\_v2 exhibit a strong recall of 99.75\%. Additionally, DenseNet169 secured the second-highest recall of 99.50\% compared to our proposed framework. While these transfer learning approaches demonstrate strong performance, XCCNet ultimately surpasses them in terms of overall effectiveness. This suggests that XCCNet's unique architecture and design choices, including contrastive learning and explainability, contribute significantly to its superior performance in pediatric pneumonia detection.

\subsubsection{Ablation Studies}

To systematically assess the impact of individual components on XCCNet's overall performance, we conducted an ablation study, analyzing the effectiveness of each proposed module and strategy (detailed in~\autoref{tab:ablation}). We specifically evaluated the contributions of the synthetic data generator, CXR enhancer, segmentator, rib-suppressor, and feature extractor. XCCNet integrates an Adversarial-based Data Augmentation unit (ADA), Contrast Limited Adaptive Histogram Equalization (CLAHE), ResUNet++, and a combination of Spatial Feature Extractor (SFx) and Contrastive-based Transformer Feature Extractor (CoTFx) modules. Notably, combining all components yielded the highest accuracy of 99.76\% and F1-score of 99.75\% for pediatric pneumonia detection, highlighting the synergistic effect of our proposed framework.

\begin{table*}[!ht] \scriptsize
\centering
\caption{Ablation Study of proposed XCCNet on Kermany Dataset}
    \begin{tabular*}{0.9\linewidth}{@{\extracolsep{\fill}}l cc cc cc cc}
    % \begin{tabular}{l cc cc cc cc}
    \toprule
    \textbf{Methods} & \textbf{Acc$^a$} (\%) & $|\nabla|$ & \textbf{ Pre$^b$} (\%) & $|\nabla|$ & \textbf{ Rec$^c$} (\%) & $|\nabla|$ & \textbf{ F1$^d$} (\%) & $|\nabla|$\\              
    \midrule
    \multicolumn{9}{l}{\textbf{Effect of Synthetic Data Generator}}\\
    \midrule
    No Augmentation &96.29 & 3.47 & 94.31 & 5.44 & 97.48 & 2.27 & 95.87 & 3.88 \\
    Traditional Augmentation &97.18 & 2.58 & 94.52 & 5.23 & 98.65 & 1.10 & 96.54 & 3.21 \\
    
    \midrule
    \multicolumn{9}{l}{\textbf{Effect of CXR Enhancer}} \\
    \midrule
    Raw CXR &96.88 & 2.88 & 94.57 & 5.18 & 94.32 & 5.43 & 94.44 & 5.31 \\
    Contrastive Enhancement &97.01 & 2.75 & 94.22 & 5.53 & 95.64 & 4.11 & 94.92 & 4.83 \\
    Histogram Equalization &97.88 & 1.88 & 96.83 & 2.92 & 97.52 & 2.23 & 97.17 & 2.58 \\
    Sharpen &94.54 & 5.22 & 96.66 & 3.09 & 96.09 & 3.66 & 96.37 & 3.38 \\
    
    \midrule
    \multicolumn{9}{l}{\textbf{Effect of Segmentator}} \\
    \midrule
    UNet &97.84 & 1.92 & 96.94 & 2.81 & 98.55 & 1.20 & 97.74 & 2.01 \\
    UNet++ &98.02 & 1.74 & 98.35 & 1.40 & 97.15 & 2.60 & 97.75 & 2.00 \\
        
    \midrule
    \multicolumn{9}{l}{\textbf{Effect of Rib Suppressor}} \\
    \midrule    
    w/o Rib Suppressor &98.59 & 1.17 & 96.83 & 2.92 & 97.48 & 2.27 & 97.15 & 2.60 \\

    \midrule
    \multicolumn{9}{l}{\textbf{Effect of Feature Extractor}} \\
    \midrule
    SFx Module &98.36 & 1.40 & 98.84 & 0.91 & 99.01 & 0.74 & 98.92 & 0.83 \\
    CoTFx Module &99.19 & 0.57 & 98.43 & 1.32 & 98.37 & 1.38 & 98.40 & 1.35 \\
    
    \midrule
    \multicolumn{9}{l}{\textbf{Explainable Contrastive-based Dilated Convolutional Network with Transformer (Ours)}} \\
    \midrule
    XCCNet &99.76 & 0.00 & 99.75 & 0.00 & 99.75 & 0.00 & 99.75 & 0\\
    \bottomrule   
    \multicolumn{9}{l}{\scriptsize $^a$ Accuracy, $^b$ Precision, $^c$ Recall, $^d$ F1-Score, w/o shorthand for ``without''}
    % \end{tabular}
    \end{tabular*}
    \label{tab:ablation}
\end{table*}

\subsubsection{Explainability Analysis of XCCCNet Prediction}

\begin{figure*}[!ht]
    \centering    
    \includegraphics[width=0.9\textwidth]{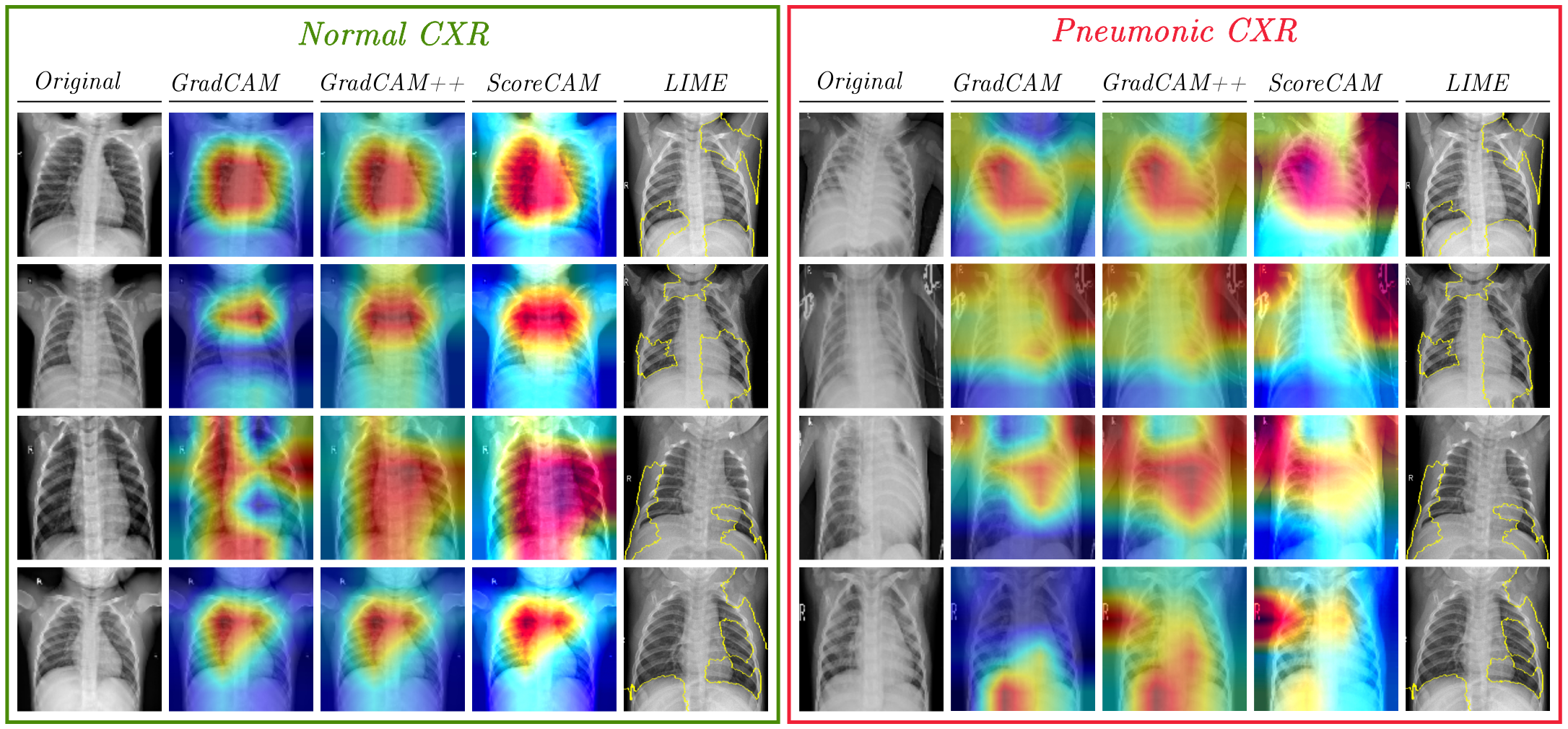}
    \caption {Explainable visualization examples using GradCAM, GradCAM++, ScoreCAM, and LIME activation maps applied to the proposed XCCNet framework. The visualizations are generated on eight chest X-rays from the Kermany Dataset, focusing on both normal and pneumonic classes.}
    \label{fig:xai}
\end{figure*}

Our comparison with state-of-the-art techniques confirmed the effectiveness of XCCNet. To investigate its behavior and feature focus further, we employed five vision-based explainability algorithms: GradCAM, GradCAM++, ScoreCAM, and Local Interpretable Model Agnostic Explanation (LIME). These algorithms pinpoint areas of potential pneumonia on chest X-rays. Visualization results in~\autoref{fig:xai} illustrate activation maps that highlight where the model focuses its attention. Darker regions indicate areas where XCCNet focuses, suggesting that these areas are more important for accurate prediction. The explainability visualizations suggest that potential pneumonia resides primarily in the middle chest, with some localization on the left or right lung. This demonstrates XCCNet's ability to focus on relevant information across different chest X-ray regions while ignoring noise and irrelevant features, contributing to its high accuracy in detecting pneumonia. By providing informed interpretations of lung pathology in the context of pneumonia, our explainability analysis can assist medical experts and radiographers in their decision-making processes.

%%%%%%%%%%%%%%% EXPERIMENTAL EVALUATIONS END %%%%%%%%%%%%%%%

\section{Discussion}
\label{sec:discussion}

\begin{figure}[!ht]
  \centering
  \includegraphics[width=0.85\linewidth]{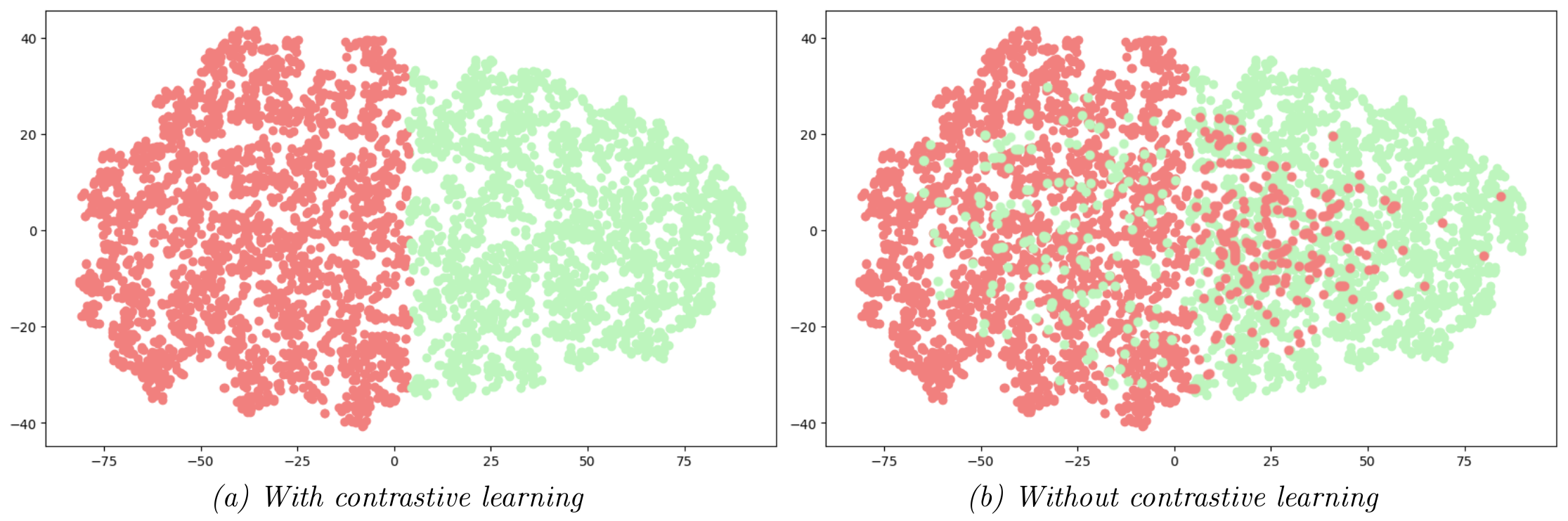}
  \caption{t-SNE visualization of data representations with and without contrastive learning. Green and red dots represent normal and pneumonia samples, respectively. (a) Feature representations obtained using contrastive learning exhibit tighter clustering and clear separation between classes. (b) Feature representations without contrastive learning show looser clustering and less distinct class boundaries.}
  \label{fig:tsne}
\end{figure}

This work introduces XCCNet, an Explainable Contrastive-based Dilated Convolutional Network with Transformer designed for pediatric pneumonia detection. XCCNet leverages the synergy of contrastive learning, transformers, and convolutional networks to effectively capture both global and spatial features from chest X-rays (CXR). We implement contrastive learning as a loss function that bridges the gap between global and spatial features, boosting CXR representation by focusing on intra-class similarity and inter-class dissimilarity. \autoref{fig:tsne} visualizes the impact of contrastive learning on data representations. We observe a significant difference in data representation when applying contrastive learning. Notably, the data points exhibit tighter clustering and enable clear linear separation of the two classes, suggesting notably that contrastive learning enhances the data's inherent structure. We evaluate XCCNet's performance and robustness against existing methods on four diverse datasets, demonstrating its superiority. Furthermore, we integrate explainability into the framework to reveal where it focuses its attention. These attention maps highlight the most influential regions by assigning weights identifying the primary contributors to pediatric pneumonia detection.

In this work, we address the challenge of utilizing low-radiation, unprocessed chest CXRs for pediatric pneumonia diagnosis. This is further compounded by the complex anatomical variability inherent in pediatric lungs, including ambiguous lung boundaries, overlapping organs, reduced lung size, the presence of foreign objects within the CXR, and potential patient positioning inconsistencies. Additionally, the skewed distribution of pediatric CXR data presents a significant obstacle for deep learning models. This skew hinders their ability to effectively learn both global and local patterns within the CXR images, ultimately impacting their diagnostic accuracy.

\begin{figure*}[!ht]
    \centering    
    \includegraphics[width=\textwidth]{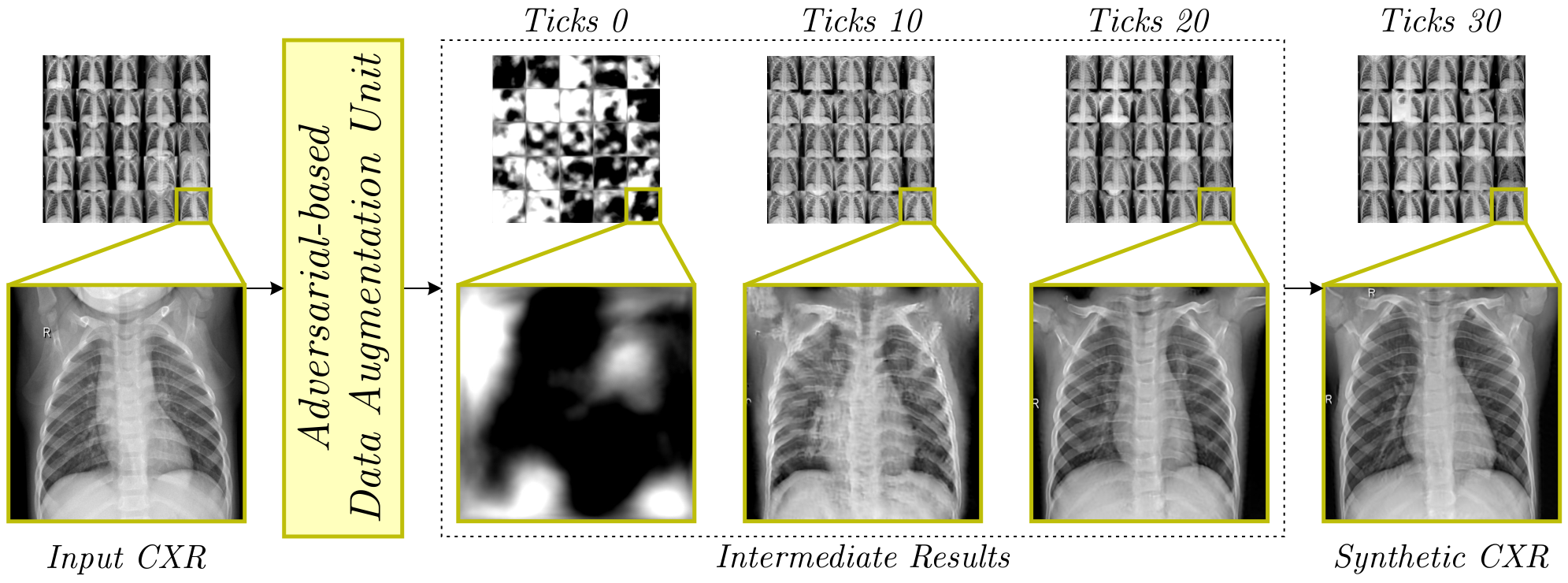}
    \caption {Synthetic CXR Samples Generated using Adversarial-based Data Augmentation Unit}
    \label{fig:adares}
\end{figure*}

In an effort to overcome the limitations associated with low-radiation, unprocessed CXR for pediatric pneumonia diagnosis, this work introduces the Chest X-ray Preprocessing (CXP) module (detailed in~\autoref{sec:cxpu}). This crucial gateway transforms raw CXRs through a series of targeted refinements. Firstly, contrast enhancement amplifies low-radiation signals, followed by segmentation to extract lung information from the CXR image meticulously. Finally, the rib suppressor selectively reduces the visibility of ribs while preserving essential internal soft tissue details. To address the skewed distribution of CXR data, we introduce a novel Adversarial-based Data Augmentation (ADA) unit, also integrated within the CXP module (refer~\autoref{sec:cxpu}). Leveraging adversarial learning principles, the ADA unit generates high-resolution synthetic CXR images, effectively mitigating the impact of data imbalance issues prevalent in medical datasets. The synthetic CXR generation process involves a series of intermediate steps known as ``ticks'', as illustrated in~\autoref{fig:adares}.

\subsection{Practical implications}

XCCNet offers a transformative approach to pediatric pneumonia diagnosis, impacting clinical practice and patient care. As a supportive second opinion, it can alleviate radiologist workload, particularly in resource-constrained settings, streamlining the diagnostic process. Furthermore, XCCNet's ability to enhance low-dose images offers the potential for reduced radiation exposure, which is especially crucial for children undergoing multiple imaging tests. This translates to improved patient safety and reduced healthcare costs. Beyond accuracy, XCCNet empowers clinicians through Explainable AI. By offering insights into the reasoning behind diagnoses, it empowers clinicians with transparency and trust in this AI-driven tool. This facilitates the acceptance and integration of XCCNet into clinical workflows. 

\subsection{Limitations}

While demonstrating XCCNet's effectiveness in discriminating between normal and pneumonia cases, our research highlights limitations. Firstly, the model currently lacks the ability to assess disease severity, such as mild, moderate, or critical stages, which is crucial for guiding treatment decisions. Additionally, our approach relies on manually labeled data, which requires significant effort and expertise, potentially hindering its scalability and generalizability. Addressing these limitations, such as incorporating severity classification and exploring semi-supervised or weakly supervised learning techniques, is essential for advancing XCCNet's development and integrating it into clinical practice.

\subsection{Directions for Future Research}

Moving forward, we aim to address XCCNet's limitations by expanding the training dataset to include chest X-rays with diverse pneumonia severities. Additionally, we plan to explore a supervised learning framework that effectively learns from a limited set of labeled CXR images, mitigating the reliance on extensive manual labeling. Furthermore, incorporating Reinforcement Learning from Human Feedback (RLHF) will accelerate XCCNet's learning curve, enabling it to grasp subtle or complex pneumonia patterns more efficiently. These advancements promise to enhance the XCCNet framework into a powerful tool for pneumonia detection and severity assessment, ultimately contributing to improved patient care and treatment outcomes.

%%%%%%%%%%%%%%% CONCLUSION START %%%%%%%%%%%%%%%
\section{Conclusion}
\label{sec:conclusion}

This paper presents the Explainable Contrastive-based Dilated Convolutional Network with Transformer (XCCNet) designed for pediatric pneumonia detection in chest X-rays (CXR). XCCNet leverages the synergy of contrastive learning, transformers, and convolutional networks to extract global and fine-grained spatial features from CXRs effectively. By leveraging contrastive learning as a loss function, XCCNet bridges the gap between global and spatial features, significantly boosting CXR representation by focusing on intra-class similarity and inter-class dissimilarity. To address the challenges of low-radiation, unprocessed CXRs, we integrate a preprocessing module that enhances the intensity and extracts the lung area, focusing analysis on the region of interest. Furthermore, a novel adversarial-based data augmentation efficiently tackles the issue of skewed CXR data distribution by generating high-quality synthetic CXR. Finally, by integrating explainability into XCCNet, we provide valuable internal insights into lung pathology, assisting medical experts and radiographers in their decision-making processes.

%%%%%%%%%%%%%%% CONCLUSION END %%%%%%%%%%%%%%%

\section*{Acknowledgment}
% DONE
We are thankful for the Young Faculty Research Catalysing Grant (YFRCG) by Indian Institute of Technology Indore for providing the resources to the project (Project ID: IITI/YFRCG/2023-24/03).

\bibliographystyle{elsarticle-num} % numbered reference
% \bibliographystyle{model2-names.bst} % named reference 
% \biboptions{authoryear}

\bibliography{refs}

\begin{thebibliography}{10}
\expandafter\ifx\csname url\endcsname\relax
  \def\url#1{\texttt{#1}}\fi
\expandafter\ifx\csname urlprefix\endcsname\relax\def\urlprefix{URL }\fi
\expandafter\ifx\csname href\endcsname\relax
  \def\href#1#2{#2} \def\path#1{#1}\fi

\bibitem{Liang2020}
G.~Liang, L.~Zheng, A transfer learning method with deep residual network for pediatric pneumonia diagnosis, Computer Methods and Programs in Biomedicine 187 (4 2020).
\newblock \href {https://doi.org/10.1016/j.cmpb.2019.06.023} {\path{doi:10.1016/j.cmpb.2019.06.023}}.

\bibitem{Fernandes2021}
V.~Fernandes, G.~B. Junior, A.~C. de~Paiva, A.~C. Silva, M.~Gattass, Bayesian convolutional neural network estimation for pediatric pneumonia detection and diagnosis, Computer Methods and Programs in Biomedicine 208 (2021).
\newblock \href {https://doi.org/10.1016/j.cmpb.2021.106259} {\path{doi:10.1016/j.cmpb.2021.106259}}.

\bibitem{Raoof2012}
S.~Raoof, D.~Feigin, A.~Sung, S.~Raoof, L.~Irugulpati, E.~C. Rosenow, Interpretation of plain chest roentgenogram (2012).
\newblock \href {https://doi.org/10.1378/chest.10-1302} {\path{doi:10.1378/chest.10-1302}}.

\bibitem{Delrue2011}
L.~Delrue, R.~Gosselin, B.~Ilsen, A.~V. Landeghem, J.~de~Mey, P.~Duyck, Difficulties in the Interpretation of Chest Radiography, Springer, 2011, Ch.~2, pp. 27--49.
\newblock \href {https://doi.org/10.1007/978-3-540-79942-9_2} {\path{doi:10.1007/978-3-540-79942-9_2}}.

\bibitem{He2019}
J.~He, S.~L. Baxter, J.~Xu, J.~Xu, X.~Zhou, K.~Zhang, The practical implementation of artificial intelligence technologies in medicine (2019).
\newblock \href {https://doi.org/10.1038/s41591-018-0307-0} {\path{doi:10.1038/s41591-018-0307-0}}.

\bibitem{Liu2024effctm}
S.~Liu, L.~Wang, W.~Yue, An efficient medical image classification network based on multi-branch cnn, token grouping transformer and mixer mlp, Applied Soft Computing 153 (2024) 111323.
\newblock \href {https://doi.org/10.1016/j.asoc.2024.111323} {\path{doi:10.1016/j.asoc.2024.111323}}.

\bibitem{Kaya2024featurefusion}
M.~Kaya, Feature fusion-based ensemble cnn learning optimization for automated detection of pediatric pneumonia, Biomedical Signal Processing and Control 87 (2024).
\newblock \href {https://doi.org/10.1016/j.bspc.2023.105472} {\path{doi:10.1016/j.bspc.2023.105472}}.

\bibitem{Kilicarslan2023}
S.~Kiliçarslan, C.~Közkurt, S.~Baş, A.~Elen, Detection and classification of pneumonia using novel superior exponential (supex) activation function in convolutional neural networks, Expert Systems with Applications 217 (2023).
\newblock \href {https://doi.org/10.1016/j.eswa.2023.119503} {\path{doi:10.1016/j.eswa.2023.119503}}.

\bibitem{Okolo2022}
G.~I. Okolo, S.~Katsigiannis, N.~Ramzan, Ievit: An enhanced vision transformer architecture for chest x-ray image classification, Computer Methods and Programs in Biomedicine 226 (2022).
\newblock \href {https://doi.org/10.1016/j.cmpb.2022.107141} {\path{doi:10.1016/j.cmpb.2022.107141}}.

\bibitem{Kaya2023}
Y.~Kaya, E.~Gürsoy, A mobilenet-based cnn model with a novel fine-tuning mechanism for covid-19 infection detection, Soft Computing 27 (2023).
\newblock \href {https://doi.org/10.1007/s00500-022-07798-y} {\path{doi:10.1007/s00500-022-07798-y}}.

\bibitem{Alharbi2022}
A.~H. Alharbi, H.~A.~H. Mahmoud, Pneumonia transfer learning deep learning model from segmented x-rays, Healthcare (Switzerland) 10 (2022).
\newblock \href {https://doi.org/10.3390/healthcare10060987} {\path{doi:10.3390/healthcare10060987}}.

\bibitem{Gupta2022}
S.~Gupta, A.~Panwar, Deep models for analysis of pneumonia infection using chest radiographs, in: Proceedings of International Conference on Information Technology and Applications: ICITA 2021, Vol. 350, 2022, pp. 263--272.
\newblock \href {https://doi.org/10.1007/978-981-16-7618-5_23} {\path{doi:10.1007/978-981-16-7618-5_23}}.

\bibitem{Pal2023}
J.~Pal, S.~Das, Prediction of Pneumonia Using Deep Convolutional Neural Network (CNN), Springer, 2023, Ch.~11, pp. 129--142.
\newblock \href {https://doi.org/10.1007/978-981-19-8742-7_11} {\path{doi:10.1007/978-981-19-8742-7_11}}.

\bibitem{Desai2022}
A.~B. Desai, D.~R. Gangodkar, B.~Pant, K.~Pant, Comparative analysis using transfer learning models vgg16, resnet 50 and xception to predict pneumonia, in: 2022 2nd International Conference on Innovative Sustainable Computational Technologies (CISCT), 2022, pp. 1--6.
\newblock \href {https://doi.org/10.1109/CISCT55310.2022.10046507} {\path{doi:10.1109/CISCT55310.2022.10046507}}.

\bibitem{Ayan2022}
E.~Ayan, B.~Karabulut, H.~M. Ünver, Diagnosis of pediatric pneumonia with ensemble of deep convolutional neural networks in chest x-ray images, Arabian Journal for Science and Engineering 47 (2022).
\newblock \href {https://doi.org/10.1007/s13369-021-06127-z} {\path{doi:10.1007/s13369-021-06127-z}}.

\bibitem{Trivedi2022}
M.~Trivedi, A.~Gupta, A lightweight deep learning architecture for the automatic detection of pneumonia using chest x-ray images, Multimedia Tools and Applications 81 (2022).
\newblock \href {https://doi.org/10.1007/s11042-021-11807-x} {\path{doi:10.1007/s11042-021-11807-x}}.

\bibitem{Chattopadhyay2022}
S.~Chattopadhyay, R.~Kundu, P.~K. Singh, S.~Mirjalili, R.~Sarkar, Pneumonia detection from lung x-ray images using local search aided sine cosine algorithm based deep feature selection method, International Journal of Intelligent Systems 37 (2022).
\newblock \href {https://doi.org/10.1002/int.22703} {\path{doi:10.1002/int.22703}}.

\bibitem{Han2024dmcnn}
Q.~Han, X.~Qian, H.~Xu, K.~Wu, L.~Meng, Z.~Qiu, T.~Weng, B.~Zhou, X.~Gao, Dm-cnn: Dynamic multi-scale convolutional neural network with uncertainty quantification for medical image classification, Computers in Biology and Medicine 168 (2024) 107758.
\newblock \href {https://doi.org/10.1016/j.compbiomed.2023.107758} {\path{doi:10.1016/j.compbiomed.2023.107758}}.

\bibitem{Rajeashwari2024deepcnn}
S.~Rajeashwari, K.~Arunesh, Enhancing pneumonia diagnosis with ensemble-modified classifier and transfer learning in deep-cnn based classification of chest radiographs, Biomedical Signal Processing and Control 93 (2024) 106130.
\newblock \href {https://doi.org/10.1016/j.bspc.2024.106130} {\path{doi:10.1016/j.bspc.2024.106130}}.

\bibitem{Yi2023dl}
R.~Yi, L.~Tang, Y.~Tian, J.~Liu, Z.~Wu, Identification and classification of pneumonia disease using a deep learning-based intelligent computational framework, Neural Computing and Applications 35 (2023).
\newblock \href {https://doi.org/10.1007/s00521-021-06102-7} {\path{doi:10.1007/s00521-021-06102-7}}.

\bibitem{Hao2024dbmvit}
Y.~Hao, C.~Zhang, X.~Li, Dbm-vit: A multiscale features fusion algorithm for health status detection in cxr / ct lungs images, Biomedical Signal Processing and Control 87 (2024).
\newblock \href {https://doi.org/10.1016/j.bspc.2023.105365} {\path{doi:10.1016/j.bspc.2023.105365}}.

\bibitem{Chen2024intercnn}
S.~Chen, S.~Ren, G.~Wang, M.~Huang, C.~Xue, Interpretable cnn-multilevel attention transformer for rapid recognition of pneumonia from chest x-ray images, IEEE Journal of Biomedical and Health Informatics 28 (2024).
\newblock \href {https://doi.org/10.1109/JBHI.2023.3247949} {\path{doi:10.1109/JBHI.2023.3247949}}.

\bibitem{Usman2022}
M.~Usman, T.~Zia, A.~Tariq, Analyzing transfer learning of vision transformers for interpreting chest radiography, Journal of Digital Imaging 35 (2022).
\newblock \href {https://doi.org/10.1007/s10278-022-00666-z} {\path{doi:10.1007/s10278-022-00666-z}}.

\bibitem{Singh2024evit}
S.~Singh, M.~Kumar, A.~Kumar, B.~K. Verma, K.~Abhishek, S.~Selvarajan, Efficient pneumonia detection using vision transformers on chest x-rays, Scientific Reports 14 (2024).
\newblock \href {https://doi.org/10.1038/s41598-024-52703-2} {\path{doi:10.1038/s41598-024-52703-2}}.

\bibitem{Mann2024hybrid}
P.~S. Mann, S.~D. Panchal, S.~Singh, G.~S. Saggu, K.~Gupta, A hybrid deep convolutional neural network model for improved diagnosis of pneumonia, Neural Computing and Applications 36 (2024) 1791--1804.
\newblock \href {https://doi.org/10.1007/s00521-023-09147-y} {\path{doi:10.1007/s00521-023-09147-y}}.

\bibitem{Gupta2023hybridcnn}
H.~Gupta, N.~Bansal, S.~Garg, H.~Mallik, A.~Prabha, J.~Yadav, A hybrid convolutional neural network model to detect covid-19 and pneumonia using chest x-ray images, International Journal of Imaging Systems and Technology 33 (2023).
\newblock \href {https://doi.org/10.1002/ima.22829} {\path{doi:10.1002/ima.22829}}.

\bibitem{raghaw2024cotconet}
C.~S. Raghaw, A.~Sharma, S.~Bansal, M.~Z.~U. Rehman, N.~Kumar, Cotconet: An optimized coupled transformer-convolutional network with an adaptive graph reconstruction for leukemia detection, Computers in Biology and Medicine 179 (2024) 108821.
\newblock \href {https://doi.org/10.1016/j.compbiomed.2023.107407} {\path{doi:10.1016/j.compbiomed.2023.107407}}.

\bibitem{Prakash2023tl}
J.~A. Prakash, C.~R. Asswin, K.~S.~D. Kumar, A.~Dora, V.~Ravi, V.~Sowmya, E.~A. Gopalakrishnan, K.~P. Soman, Transfer learning approach for pediatric pneumonia diagnosis using channel attention deep cnn architectures, Engineering Applications of Artificial Intelligence 123 (2023).
\newblock \href {https://doi.org/10.1016/j.engappai.2023.106416} {\path{doi:10.1016/j.engappai.2023.106416}}.

\bibitem{Dai2024xai}
L.~Dai, X.~Yang, H.~Li, X.~Zhao, L.~Lin, Y.~Jiang, Y.~Wang, Z.~Li, H.~Shen, A clinically actionable and explainable real-time risk assessment framework for stroke-associated pneumonia, Artificial Intelligence in Medicine 149 (2024) 102772.
\newblock \href {https://doi.org/10.1016/j.artmed.2024.102772} {\path{doi:10.1016/j.artmed.2024.102772}}.

\bibitem{Hasan2023fpcnn}
M.~M. Hasan, M.~M. Hossain, M.~M. Rahman, A.~K. Azad, S.~A. Alyami, M.~A. Moni, Fp-cnn: Fuzzy pooling-based convolutional neural network for lung ultrasound image classification with explainable ai, Computers in Biology and Medicine 165 (2023).
\newblock \href {https://doi.org/10.1016/j.compbiomed.2023.107407} {\path{doi:10.1016/j.compbiomed.2023.107407}}.

\bibitem{Ukwuoma2023xet}
C.~C. Ukwuoma, Z.~Qin, M.~B.~B. Heyat, F.~Akhtar, O.~Bamisile, A.~Y. Muaad, D.~Addo, M.~A. Al-antari, A hybrid explainable ensemble transformer encoder for pneumonia identification from chest x-ray images, Journal of Advanced Research 48 (2023).
\newblock \href {https://doi.org/10.1016/j.jare.2022.08.021} {\path{doi:10.1016/j.jare.2022.08.021}}.

\bibitem{Yang2022xai}
Y.~Yang, G.~Mei, F.~Piccialli, A deep learning approach considering image background for pneumonia identification using explainable ai (xai), IEEE/ACM Transactions on Computational Biology and Bioinformatics (2022).
\newblock \href {https://doi.org/10.1109/TCBB.2022.3190265} {\path{doi:10.1109/TCBB.2022.3190265}}.

\bibitem{Hroub2024xdl}
N.~A. Hroub, A.~N. Alsannaa, M.~Alowaifeer, M.~Alfarraj, E.~Okafor, Explainable deep learning diagnostic system for prediction of lung disease from medical images, Computers in Biology and Medicine 170 (2024) 108012.
\newblock \href {https://doi.org/10.1016/j.compbiomed.2024.108012} {\path{doi:10.1016/j.compbiomed.2024.108012}}.

\bibitem{Chen2022ganreview}
Y.~Chen, X.~H. Yang, Z.~Wei, A.~A. Heidari, N.~Zheng, Z.~Li, H.~Chen, H.~Hu, Q.~Zhou, Q.~Guan, Generative adversarial networks in medical image augmentation: A review, Computers in Biology and Medicine 144 (2022).
\newblock \href {https://doi.org/10.1016/j.compbiomed.2022.105382} {\path{doi:10.1016/j.compbiomed.2022.105382}}.

\bibitem{Xu2024cdagan}
Z.~Xu, J.~Tang, C.~Qi, D.~Yao, C.~Liu, Y.~Zhan, T.~Lukasiewicz, Cross-domain attention-guided generative data augmentation for medical image analysis with limited data, Computers in Biology and Medicine 168 (2024) 107744.
\newblock \href {https://doi.org/10.1016/j.compbiomed.2023.107744} {\path{doi:10.1016/j.compbiomed.2023.107744}}.

\bibitem{Shamrat2023clahe}
F.~J.~M. Shamrat, S.~Azam, A.~Karim, K.~Ahmed, F.~M. Bui, F.~D. Boer, High-precision multiclass classification of lung disease through customized mobilenetv2 from chest x-ray images, Computers in Biology and Medicine 155 (2023).
\newblock \href {https://doi.org/10.1016/j.compbiomed.2023.106646} {\path{doi:10.1016/j.compbiomed.2023.106646}}.

\bibitem{Rifai2024imgenhance}
A.~M. Rifai, S.~Raharjo, E.~Utami, D.~Ariatmanto, Analysis for diagnosis of pneumonia symptoms using chest x-ray based on mobilenetv2 models with image enhancement using white balance and contrast limited adaptive histogram equalization (clahe), Biomedical Signal Processing and Control 90 (2024) 105857.
\newblock \href {https://doi.org/10.1016/j.bspc.2023.105857} {\path{doi:10.1016/j.bspc.2023.105857}}.

\bibitem{Rajaraman2021}
S.~Rajaraman, G.~Zamzmi, L.~Folio, P.~Alderson, S.~Antani, Chest x-ray bone suppression for improving classification of tuberculosis-consistent findings, Diagnostics 11 (2021).
\newblock \href {https://doi.org/10.3390/diagnostics11050840} {\path{doi:10.3390/diagnostics11050840}}.

\bibitem{Radford2021}
A.~Radford, J.~W. Kim, C.~Hallacy, A.~Ramesh, G.~Goh, S.~Agarwal, G.~Sastry, A.~Askell, P.~Mishkin, J.~Clark, G.~Krueger, I.~Sutskever, Learning transferable visual models from natural language supervision, in: International conference on machine learning, Vol. 139, 2021, pp. 8748--8763.

\bibitem{Wang2022cltranscnn}
X.~Wang, S.~Yang, J.~Zhang, M.~Wang, J.~Zhang, W.~Yang, J.~Huang, X.~Han, Transformer-based unsupervised contrastive learning for histopathological image classification, Medical Image Analysis 81 (2022).
\newblock \href {https://doi.org/10.1016/j.media.2022.102559} {\path{doi:10.1016/j.media.2022.102559}}.

\bibitem{Kermany2018}
D.~S. Kermany, M.~Goldbaum, W.~Cai, C.~C. Valentim, H.~Liang, S.~L. Baxter, A.~McKeown, G.~Yang, X.~Wu, F.~Yan, J.~Dong, M.~K. Prasadha, J.~Pei, M.~Ting, J.~Zhu, C.~Li, S.~Hewett, J.~Dong, I.~Ziyar, A.~Shi, R.~Zhang, L.~Zheng, R.~Hou, W.~Shi, X.~Fu, Y.~Duan, V.~A. Huu, C.~Wen, E.~D. Zhang, C.~L. Zhang, O.~Li, X.~Wang, M.~A. Singer, X.~Sun, J.~Xu, A.~Tafreshi, M.~A. Lewis, H.~Xia, K.~Zhang, Identifying medical diagnoses and treatable diseases by image-based deep learning, Cell 172 (2018).
\newblock \href {https://doi.org/10.1016/j.cell.2018.02.010} {\path{doi:10.1016/j.cell.2018.02.010}}.

\bibitem{Pham2023vindr}
H.~H. Pham, N.~H. Nguyen, T.~T. Tran, T.~N. Nguyen, H.~Q. Nguyen, Pedicxr: An open, large-scale chest radiograph dataset for interpretation of common thoracic diseases in children, Scientific Data 10 (2023).
\newblock \href {https://doi.org/10.1038/s41597-023-02102-5} {\path{doi:10.1038/s41597-023-02102-5}}.

\bibitem{Wang2017nihpediatric}
X.~Wang, Y.~Peng, L.~Lu, Z.~Lu, M.~Bagheri, R.~M. Summers, Chestx-ray8: Hospital-scale chest x-ray database and benchmarks on weakly-supervised classification and localization of common thorax diseases, in: Proceedings - 30th IEEE Conference on Computer Vision and Pattern Recognition, CVPR 2017, Vol. 2017-January, 2017, pp. 2097--2106.
\newblock \href {https://doi.org/10.1109/CVPR.2017.369} {\path{doi:10.1109/CVPR.2017.369}}.

\end{thebibliography}
\end{document}